\documentclass{emulateapj}

\usepackage{verbatim}
\usepackage{natbib}
\usepackage{amsmath}
\usepackage{amsbsy}
\usepackage{lscape}
\usepackage{epstopdf}

\slugcomment{Published in ApJ, 681, 831}
\shorttitle{Intergalactic Opacity at $2\leq z \leq 4.2$}
\shortauthors{Faucher-Gigu\`ere et al.}

\newcommand{\Lya}{\mbox{Ly$\alpha$}}
\newcommand{\Gbkg}{\mbox{$\Gamma^{bkg}$}}

\begin{document}

\title{A Direct Precision Measurement of the Intergalactic Lyman-$\alpha$ Opacity at $2 \leq z \leq 4.2$\altaffilmark{1,2}}

\author{Claude-Andr\'e Faucher-Gigu\`ere\altaffilmark{3}, Jason X. Prochaska\altaffilmark{4}, Adam Lidz\altaffilmark{3}, Lars Hernquist\altaffilmark{3}, Matias Zaldarriaga\altaffilmark{3,5}}
\altaffiltext{1}{Based, in part, on data obtained at the W.M. Keck
Observatory, which is operated as a scientific partnership among the
California Institute of Technology, the University of California and the
National Aeronautics and Space Administration. The Observatory was made
possible by the generous financial support of the W.~M. Keck Foundation.}
\altaffiltext{2}{Some of the data analyzed in this work were gathered with the 6.5 meter Magellan Telescopes located at Las Campanas Observatory, Chile.}
\altaffiltext{3}{Department of Astronomy, Harvard University, Cambridge, MA, 02138, USA; cgiguere@cfa.harvard.edu.}
\altaffiltext{4}{Department of Astronomy and Astrophysics, UCO/Lick Observatory; University of California, 1156 High Street, Santa Cruz, CA, 95064, USA.}
\altaffiltext{5}{Jefferson Physical Laboratory, Harvard University, Cambridge, MA, 02138, USA.}

\begin{abstract}
We directly measure the evolution of the intergalactic Lyman-$\alpha$ effective optical depth, $\tau_{\rm eff}$, over the redshift range $2 \leq z \leq 4.2$ from a sample of 86 high-resolution, high-signal-to-noise quasar spectra obtained with the ESI and HIRES spectrographs on Keck, and with the MIKE spectrograph on Magellan. 
This represents an improvement over previous analyses of the \Lya~forest from high-resolution spectra in this redshift interval of a factor of two in the size of the data set alone. 
We pay particular attention to robust error estimation and extensively test for systematic effects.
We find that our estimates of the quasar continuum levels in the \Lya~forest obtained by spline fitting are systematically biased low, with the magnitude of the bias increasing with redshift, but that this bias can be accounted for using mock spectra.
The mean fractional error $\langle \Delta C/C_{\rm true} \rangle$ is $<1\%$ at $z=2$, 4\% at $z=3$, and 12\% at $z=4$.
Previous measurements of $\tau_{\rm eff}$ at $z\gtrsim3$ based on directly fitting the quasar continua in the \Lya~forest, which have generally neglected this effect, are therefore likely biased low.
We provide estimates of the level of absorption arising from metals in the \Lya~forest based on both direct and statistical metal removal results in the literature, finding that this contribution is $\approx6-9\%$ at $z=3$ and decreases monotonically with redshift.
The high precision of our measurement, attaining 3\% in redshift bins of width $\Delta z=$0.2 around $z=3$, indicates significant departures from the best-fit power-law redshift evolution ($\tau_{\rm eff}=0.0018(1+z)^{3.92}$, when metals are left in), particularly near $z=3.2$.
The observed downward departure is statistically consistent with a similar feature detected in a precision statistical measurement using Sloan Digital Sky Survey spectra by Bernardi and coworkers using an independent approach. 
\end{abstract}


\section{INTRODUCTION}
\label{introduction}
The evolution of the intergalactic medium (IGM) as traced by the Lyman-$\alpha$ (\Lya) forest provides a powerful record of the thermal and radiative history of the Universe.
This power owes to our ability to measure the \Lya~opacity of the
IGM as a function of redshift, from $z=0$ to $z\gtrsim6$, with high precision from quasar spectra \citep{1993ApJ...414...64P, 1997ApJ...489....7R, 1999ApJ...525L...5S, 2001ApJ...549L..11M, 2003MNRAS.342.1205M, 2004ApJ...617....1T, 2004AJ....127.2598S, 2005MNRAS.357.1178B, 2005MNRAS.360.1373K, 2005MNRAS.361...70J, 2006AJ....132..117F, 2007ApJ...662...72B, 2007astro.ph..3306B}, as well as to the relatively simple physics of the \Lya~forest.
In fact, fully and pseudo-hydrodynamical cosmological simulations, in which the forest arises from absorption by smooth density fluctuations imposed on the warm photoionized IGM as a natural consequence of hierarchical structure formation within cold dark matter models \citep{1994ApJ...437L...9C, 1995ApJ...453L..57Z, 1996ApJ...457L..51H, 1996ApJ...457L..57K, 1996ApJ...471..582M, 1998MNRAS.301..478T, 1999ApJ...511..521D}, have been remarkably successful at reproducing the properties of the absorption observed in high-resolution, high signal-to-noise quasar spectra \citep[e.g.,][]{1996ApJ...472..509L, 1997ApJ...484..672K, 2002MNRAS.335..555K}.
This synergy between theory and observations make the \Lya~forest a particularly compelling probe of the diffuse Universe.

A number of observational results suggest changes in the thermal properties of the IGM over the redshift range probed by the \Lya~forest, in particular near $z=3$.
These results fall into four classes: the mean \Lya~opacity, the widths of the \Lya~absorption lines, the strength of absorption by metallic transitions, and ultra-violet observations of the HeII forest.
\cite{2003AJ....125...32B} first claimed a statistical detection of a feature at $z\approx3.2$ in the evolution of the \Lya~forest effective optical depth from an analysis of 1061 quasar spectra from the Sloan Digital Sky Survey (SDSS; York et al. 2000\nocite{2000AJ....120.1579Y}).
These authors interpreted the observed reduction in the \Lya~effective optical depth near $z=3.2$ as evidence for the reionization of HeII \citep{2002ApJ...574L.111T}.
\cite{2000ApJ...534...41R} and \cite{2000MNRAS.318..817S} both found evidence from the Doppler widths of \Lya~absorption lines for an upward jump in the temperature of the IGM and a change of its equation of state to nearly isothermal near $z=3$.
\cite{1996AJ....112..335S} and \cite{1998AJ....115.2184S} have argued for a sudden hardening of the ionizing background near $z=3$, consistent with the IGM becoming optically thin to HeII ionizing photons, based on the observed evolution of the CIV/SiIV metal-line ratios.
Using information provided by additional metal systems, \cite{2007A&A...461..893A} argued that the IGM is optically thin to HeII ionizing photons at $z\lesssim1.8$ and measured an increasing HeII \Lya~effective optical depth from $z=2.4$ to $z=2.9$.
\cite{2003A&A...402..487V} used the abundance of ArI, which is particularly sensitive to hard photons, in damped \Lya~systems to also argue for a hardening of the ionizing background around $z=3$.   
Finally, patchy HeII absorption suggestive of the end of HeII reionization has been observed in ultra-violet spectra of quasars HE 2347-4342 \citep[$z=2.885$]{1997A&A...327..890R, 2002ApJ...564..542S} and Q0320-003 \citep[$z=3.286$]{1994Natur.370...35J, 1997AJ....113.1495H, 2000ApJ...534...69H}.
\cite{2006A&A...455...91F} interpret the HeII forest observed in a FUSE spectrum of HS 1700+6416 ($z=2.72$) as arising from residual HeII in its post-reionization epoch, while HST/STIS observations of QSO 1157+3143 ($z=3$) suggest that HeII reionization may not be complete between $z=2.77$ and $z=2.97$ \citep[][]{2005A&A...442...63R}.

Each of these observational results, however, has caveats.
\cite{2003AJ....125...32B} could not estimate the \Lya~optical depth from individual spectra, owing to the low resolution and signal-to-noise of the SDSS spectra, and instead employed a statistical method.
While this approach is in principle perfectly valid, the procedure is complex and it is difficult to ensure that it is free of systematic effects.
Moreover, \cite{2005ApJ...635..761M} also estimated the evolution of the mean transmission in the \Lya~forest from SDSS data, using different methodologies, and did not confirm the \cite{2003AJ....125...32B} feature, although they caution that their error estimates may not be accurate.
The \cite{2000ApJ...534...41R} and \cite{2000MNRAS.318..817S} temperature and equation of state measurements have large error bars and the $z\sim3$ temperature jump they find is not corroborated by the analyses of \cite{2001ApJ...562...52M} and \cite{2001ApJ...557..519Z}.
Both the \cite{2001ApJ...562...52M} and \cite{2001ApJ...557..519Z} results are consistent with an isothermal equation of state at $z\sim3$, though the evidence is weak.
Contrary to \cite{1996AJ....112..335S} and \cite{1998AJ....115.2184S}, \cite{2002A&A...383..747K}, \cite{2003astro.ph..7557B} and \cite{2004ApJ...602...38A} have not found evidence for an abrupt change in the CIV/SiIV ratio with redshift.
The $z\sim3$ observations of the HeII forest are too scarce to draw definitive conclusions on HeII reionization, especially if it is inhomogeneous.
Moreover, absorption of HeII \Lya~photons is exponentially sensitive to the HeII abundance and thus saturates at relatively low densities, making it impossible to use the HeII \Lya~forest to probe the heart of HeII reionization.

On the theoretical side, the rise of quasar luminosity function at $z\sim3$ \citep[e.g.,][]{2007ApJ...654..731H} is generally thought to go hand-in-hand with HeII reionization \citep[][]{2002MNRAS.332..601S, 2003ApJ...586..693W}, since stars are expected to emit too few photons hard enough (54.4 eV) to ionize HeII.
In comparison to the epoch of HI reionization, which has recently been the focus of detailed theoretical work aimed at predicting its timing, extent, morphology, and observational signatures \citep[e.g.,][]{2004ApJ...608..622Z, 2004ApJ...613....1F, 2006astro.ph.12406T, 2007ApJ...654...12Z, 2007arXiv0704.2239M, 2007MNRAS.377.1043M}, HeII reionization has received relatively little attention
\citep[for pioneering work, see][]{1997ApJ...488..532C, 2002MNRAS.332..601S, 2002ApJ...574L.111T}.
Yet, whereas it is unclear whether we have begun to probe the tail of HI reionization \citep[for a review of the observational constraints on HI reionization, see][]{2006ARA&A..44..415F}, or will be able to do so in the near future, the observational lines of evidence outlined above make a tantalizing case that we may be able to infer the details of HeII right now.
As a phase transition involving 25\% of the intergalactic baryonic mass and, possibly, the result of quasar feedback on the IGM, HeII reionization is obviously interesting in its own right.
However, it is also an excellent laboratory to test the theoretical framework that has been developed to model HI reionization.
Indeed, much of the same physics underlies the reionization of both HI and HeII.
In particular, it should be relatively simple to adapt the radiative transfer codes developed to simulate HI reionization to treat HeII.

Two lines of attack must be followed to understand HeII reionization.
First, the caveats on the observational results make it clear that further observational work is required to clarify the evolution of the thermal and ionization state of the IGM over the redshift range most likely to be affected by HeII reionization, $2\lesssim z \lesssim4$.
Second, in the absence of large space-based telescopes with ultra-violet spectroscopic capabilities to survey the HeII forest, the most promising avenue for probing HeII reionization appears to be in its indirect signatures carried by intergalactic hydrogen and metals.
In order to draw robust and instructive inferences about HeII reionization, theoretical work is thus required to quantitatively calculate these potential signatures.

Of course, regardless of whether HeII reionization is occurring, it is still of great interest to study the evolution of the IGM at $2\lesssim z \lesssim4$.
The mean transmission of the \Lya~forest, for example, probably provides the best constraints on the hydrogen ionizing background and its evolution\footnote{In \cite{2007astro.ph..1042F}, we showed that the existing measurements using the proximity effect are likely to biased high owing to quasars residing in overdense regions of the Universe. Observational difficulties, such as uncertainties on quasar systematic redshifts, and temporal variability of quasar luminosities, also make it challenging to obtain robust proximity effect measurements.} \citep{1997ApJ...489....7R, 1999ApJ...525L...5S, 2001ApJ...549L..11M, 2003MNRAS.342.1205M, 2004ApJ...617....1T, 2005MNRAS.357.1178B, 2005MNRAS.360.1373K, 2005MNRAS.361...70J}.
Indeed, the \Lya~optical depth at any point in the IGM is inversely proportional to the photoionization rate.
As the ionizing background is an integral over all cosmic sources of radiation, it provides a crucial tracer of star formation and quasar activity in the Universe.
The thermal state of the IGM at these redshifts should also bear relic signatures of HI reionization and provide constraints on that epoch \citep[][]{2002ApJ...567L.103T, 2003ApJ...596....9H}.
Moreover, the mean transmission $\langle F \rangle$ of the IGM is a very important quantity in the determination of cosmological parameters, in particular the matter power spectrum, from the \Lya~forest \citep[][]{1998ApJ...495...44C, 2002ApJ...581...20C, 2005ApJ...635..761M, 2006ApJ...638...27L}. \cite{2003MNRAS.342L..79S} showed, for example, that increasing $\langle F \rangle(z)$ by $5\%$ at the mean redshift 2.72 changes the best-fit power spectrum amplitude by almost a factor of 2, and its slope by 0.2, at the pivot point $k_{p}=0.03$ s km$^{-1}$ \citep[see, however,][who bypass the use of an external constraint on $\langle F \rangle$ in their estimation of the linear theory power spectrum from SDSS data]{2005ApJ...635..761M}.

In this paper, we make an important step in the observational direction by presenting a direct precision measurement of the effective \Lya~optical depth and its evolution over the redshift range $2\leq z \leq4.2$ from a large sample of 86 high-resolution, high signal-to-noise quasar spectra obtained obtained with the ESI and HIRES spectrographs on Keck and with MIKE on Magellan.
Our focus here is on the presentation of model-independent observational results with robust uncertainty estimates.
As such, we provide detailed numerical tables and fitting formulae, and leave theoretical interpretations to separate work.

We describe our data set in \S \ref{data set}.
Our measurement of the \Lya~effective optical depth and its statistical error analysis are presented in \S \ref{effective optical depth measurement}.
We also present estimates of the bias arising from the difficulty of estimating quasar continua at high redshifts and of the level of absorption from metals in the \Lya~forest in this section.
The systematic uncertainty on the continuum and metal corrections is estimated in \S \ref{systematic errors}.
In \S \ref{fits and comparisons}, we fit simple functions to our measurement, corrected for metal absorption and not, and compare our results with previous work.
We conclude in \S \ref{discussion} by outlining the physical effects that may give rise to the feature we find evidence for near $z=3.2$ and their degeneracies, leaving theoretical interpretation for future work.
A detailed appendix describes a series of tests for systematic errors.

Throughout this paper, we assume a cosmology with $(\Omega_{m},~\Omega_{b},~\Omega_{\Lambda},~h,~\sigma_{8})=(0.27,~0.046,~0.73,~0.7,~0.8)$, close to the parameters inferred from the \emph{Wilkinson Microwave Anisotropy Probe} three-year data \citep{2007ApJS..170..377S}. 

\begin{figure}[ht]
\begin{center}
\includegraphics[width=0.45\textwidth]{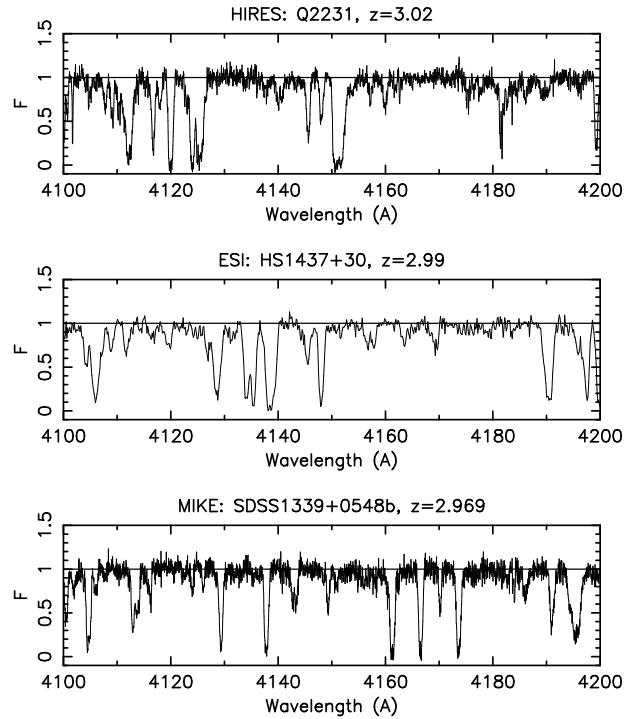}
\end{center}
\caption{Examples of quasar spectra used in this study.
The top panel shows a spectrum obtained with Keck/HIRES, the middle panel a spectrum obtained with Keck/ESI, and the bottom panel one obtained with Magellan/MIKE.
The spectra were normalized by dividing them by the estimated continuum level ($F\equiv1$) (see \S \ref{continuum estimation}), indicated by the thin horizontal line.
The ESI spectrum has lower resolution and fewer pixels.
}
\label{example spectra}
\end{figure}
\begin{figure}[ht]
\begin{center}
\includegraphics[width=0.45\textwidth]{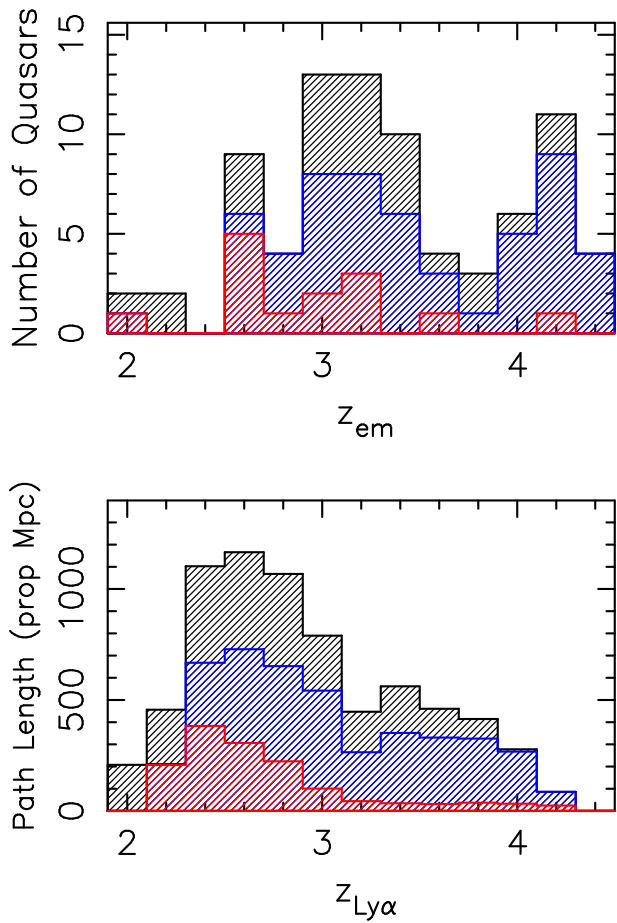}
\end{center}
\caption{Histograms showing the redshift distribution of the quasars (top panel) used in this study and the corresponding proper path length distribution of \Lya~absorption (bottom panel).
In each panel, the red histogram corresponds to HIRES data only, the blue one to HIRES+ESI data, and the black one to the complete data set (HIRES+ESI+MIKE).
}
\label{z hists}
\end{figure}
\section{DATA SET}
\label{data set}
Our data set consists of quasar spectra obtained with the High Resolution Echelle Spectrometer \citep[HIRES]{1994SPIE.2198..362V} and Spectrograph and Imager \citep[ESI]{2002PASP..114..851S} spectrographs on the Keck telescopes and with the Magellan Inamori Kyocera Echelle \citep[MIKE]{2003SPIE.4841.1694B} spectrograph on Magellan.

For nearly all observations, the HIRES spectra were acquired using either a $0.8''$ or $1.1''$ wide decker ($FWHM\approx6$ and 8 km s$^{-1}$, respectively) and the ESI observations were carried out with the $0.5''$ or $0.75''$ slit ($FWHM\approx33$ and 44 km s$^{-1}$, respectively).
In general, the signal-to-noise ratio ($S/N$) of these spectra is $\gtrsim15$ per pixel (2 km s$^{-1}$ pix$^{-1}$ for HIRES and 11 km s$^{-1}$ pix$^{-1}$ for ESI).
Most of the ESI and HIRES spectra were originally obtained to study damped \Lya~absorbers (DLAs).
A subset of these has been publicly released to the community by \cite{2007ApJS..171...29P} and were previously summarized in \cite{1999ApJS..121..369P}, \cite{2001ApJ...552...99P}, and \cite{2003ApJS..147..227P}.
A few other ESI spectra were analyzed in \cite{2007ApJ...656..666O}.
This sample is supplemented by a number of new ESI spectra also selected for DLA studies (Prochaska et al., in prep.).
Seven other HIRES spectra were provided by M. Rauch and W. Sargent \citep[e.g.][]{1997ApJ...489....7R}.

The MIKE spectrograph has a dichroic optical element splitting the beam into blue and red arms, with $FWHM\approx11$ km s$^{-1}$ and 14 km s$^{-1}$ for the blue and red sides, respectively, for a $1.0''$ slit.
Our MIKE spectra have $S/N\gtrsim10$ per pixel and were analyzed for super Lyman-limit systems (SLLS) in \cite{2007ApJ...656..666O}.
Figure \ref{example spectra} shows examples of quasar spectra obtained with each instrument used in this study.
Tables \ref{qsotab_HIRES}, \ref{qsotab_ESI}, and \ref{qsotab_MIKE} lists the quasars in the HIRES (16), ESI (44), and MIKE (26) data sets, respectively, with their redshifts.
Figure \ref{z hists} illustrates the quasar redshift distributions, as well as the corresponding path lengths of \Lya~absorption, in histogram form. 
\begin{deluxetable}{lccl}
\tablewidth{0pc}
\tablecaption{HIRES Quasar Sample\label{qsotab_HIRES}}
\tabletypesize{\footnotesize}
\tablehead{\colhead{Name} &\colhead{$z_{em}$}}
\startdata
Q0000-26\tablenotemark{a} & 4.11 \\
BR0019-15 & 4.53 \\
Q0336-01 & 3.20 \\
Q1107+49\tablenotemark{a} & 3.00 \\
Q1209+09 & 3.30 \\
Q1210+17 & 2.54 \\
Q1215+33 & 2.61 \\
Q1422+23\tablenotemark{a} & 3.62 \\ 
Q1425+60 & 3.20 \\
Q1442+29\tablenotemark{a} & 2.67 \\
Q1759+75 & 3.05 \\
Q2206-19 & 2.56 \\
Q2237-06\tablenotemark{a} & 4.55 \\
Q2231-002 & 3.02 \\
Q2343+12\tablenotemark{a} & 2.52 \\
Q2359-02 & 2.80 \\
\enddata
\tablenotetext{a}{Provided by M. Rauch and W. Sargent \citep[e.g.,][]{1997ApJ...489....7R}.}
\end{deluxetable}

\begin{deluxetable}{lccl}
\tablewidth{0pc}
\tablecaption{ESI Quasar Sample\label{qsotab_ESI}}
\tabletypesize{\footnotesize}
\tablehead{\colhead{Name} &\colhead{$z_{em}$}}
\startdata
PSS0007+2417 & 4.05 \\
SDSS0013+1358 & 3.58 \\
PX0034+16 & 4.29 \\
Q0112-30 & 2.99 \\
SDSS0127-00 & 4.06 \\
PSS0134+3307 & 4.52 \\
SDSS0139-0824 & 3.02 \\
SDSS0142+0023 & 3.39 \\
PSS0209+0517 & 4.17 \\
SDSS0225+0054 & 2.97 \\
BRJ0426-2202 & 4.32 \\
PSS0808+52 & 4.45 \\
SDSS0814+5029 & 3.88 \\
Q0821+31 & 2.61 \\
SDSS0844+5153 & 3.20 \\
SDSS0912+5621 & 3.00 \\
SDSS0912-0047 & 2.86 \\
Q0930+28 & 3.42 \\
BQ1021+30 & 3.12 \\
CTQ460 & 3.13 \\
HS1132+22 & 2.89 \\
SDSS1155+0530 & 3.48 \\
PSS1159+13 & 4.07 \\
PSS1248+31 & 4.35 \\
PSS1253-02 & 4.01 \\
SDSS1410+5111 & 3.21 \\
PSS1432+39 & 4.28 \\
HS1437+30 & 2.99 \\
PSS1506+52 & 4.18 \\
PSS1535+2943 & 3.99 \\
SDSS1610+4724 & 3.22 \\
PSS1715+3809 & 4.52 \\
PSS1723+2243 & 4.52 \\
PSS1802+5616 & 4.18 \\
FJ2129+00 & 2.96 \\
PSS2155+1358 & 4.26 \\
Q2223+20 & 3.56 \\
SDSS2238+0016 & 3.47 \\
PSS2241+1352 & 4.44 \\
SDSS2315+1456 & 3.39 \\
PSS2323+2758 & 4.18 \\
FJ2334-09 & 3.33 \\
PSS2344+0342 & 4.24 \\
SDSS2350-00 & 3.01 \\
\enddata
\end{deluxetable}

\begin{deluxetable}{lccl}
\tablewidth{0pc}
\tablecaption{MIKE Quasar Sample\label{qsotab_MIKE}}
\tabletypesize{\footnotesize}
\tablehead{\colhead{Name} &\colhead{$z_{em}$}}
\startdata
Q0101-304 & 3.14 \\
CTS0291 & 2.55 \\
HE0340-2612 & 3.08 \\
SDSS0912+0547 & 3.25 \\
SDSS0942+0422 & 3.27 \\
SDSS0949+0355 & 4.10 \\
SDSS1025+0452 & 3.24 \\
Q1100-264 & 2.14 \\
SDSS1110+0244 & 4.15 \\
Q1224-0812 & 2.14 \\
SDSS1249-0159 & 3.66 \\ 
SDSS1339+0548 & 2.97 \\
HE1347-2457 & 2.60 \\
Q1358+1154 & 2.58 \\
SDSS1402+0146 & 4.19 \\
SDSS1429-0145 & 3.42 \\
Q1456-1938 & 3.16 \\
SDSS1621-0042 & 3.70 \\
PKS2000-330 & 3.78 \\
Q2044-1650 & 1.94 \\
Q2126-158 & 3.28 \\
HE2156-4020 & 2.53 \\
HE2215-6206 & 3.32 \\
HE2314-3405 & 2.94 \\
HE2348-1444 & 2.93 \\
HE2355-5457 & 2.93 \\
\enddata
\end{deluxetable}

\section{MEASUREMENT OF THE EFFECTIVE \Lya~OPTICAL DEPTH}
\label{effective optical depth measurement}
We now measure the \Lya~opacity of the IGM as a function of redshift from the data set described in \S \ref{data set}.
We begin by defining the quantities of interest and describing our methodology, and present the measurement itself, in \S \ref{definitions}.
Masks which are applied to the data are described in \S \ref{masks}.
We discuss the calculation of error bars in \S \ref{error bars}.
In \S \ref{continuum estimation}, we test for systematic effects in estimating the continuum level of the quasar spectra, and calculate the redshift-dependent bias arising from the increasing level of absorption with increasing redshift, which we use to correct our measurement.
We estimate the level of absorption in the \Lya~forest arising from metals in \S \ref{metal absorption}.

\subsection{Definitions and Methodology}
\label{definitions}
Let $F_{\rm abs}(\lambda)$ be a quasar's absolute flux and $C(\lambda)$ be its continuum (unabsorbed) level as a function of observed wavelength $\lambda$.
The corresponding redshift of \Lya~absorption is just $z_{\Lya} \equiv \lambda/\lambda_{\Lya}-1$, where $\lambda_{\Lya}=1216$ \AA.
Dropping the \Lya~subscript on $z$, we define the transmission, or normalized flux, as
\begin{equation}
F(z)\equiv \frac{F_{\rm abs}(z)}{C(z)}.
\end{equation}
Let $\langle F \rangle$ be the ensemble average over quasars and further define the effective optical depth as
\begin{equation}
\label{tau eff}
\tau_{\rm eff}(z) \equiv -\ln{\left[\langle F \rangle(z)\right]}
\end{equation}
\citep[e.g.,][]{1997ApJ...489....7R}.
This quantity, being a function of $\langle F \rangle$ only, has the convenient property of being independent of noise, provided the noise fluctuations have a symmetric distribution about zero and sufficiently many independent pixels are averaged over.
Both conditions are satisfied for our large sample.
Moreover, it is standard in the literature, which will allow us to directly compare our results with published measurements.

The measurement consists of the following steps:
\begin{enumerate}
\item Estimate the continuum level in the \Lya~forest of each quasar spectrum and calculate the corresponding $F$;
\item apply masks to the data: quasar proximity regions, higher-order Lyman series forests, SLLS, DLAs, and associated metal lines;
\item bin the remaining \Lya~forest pixels in redshift intervals;
\item for each redshift bin, average the pixels to estimate $\langle F \rangle$ and calculate $\tau_{\rm eff}$ using Equation (\ref{tau eff});
\item correct this ``raw'' measurement for redshift-dependent continuum estimation error and, optionally, for metal absorption.
\end{enumerate}

\begin{figure*}[ht]
\begin{center}
\includegraphics[width=0.80\textwidth]{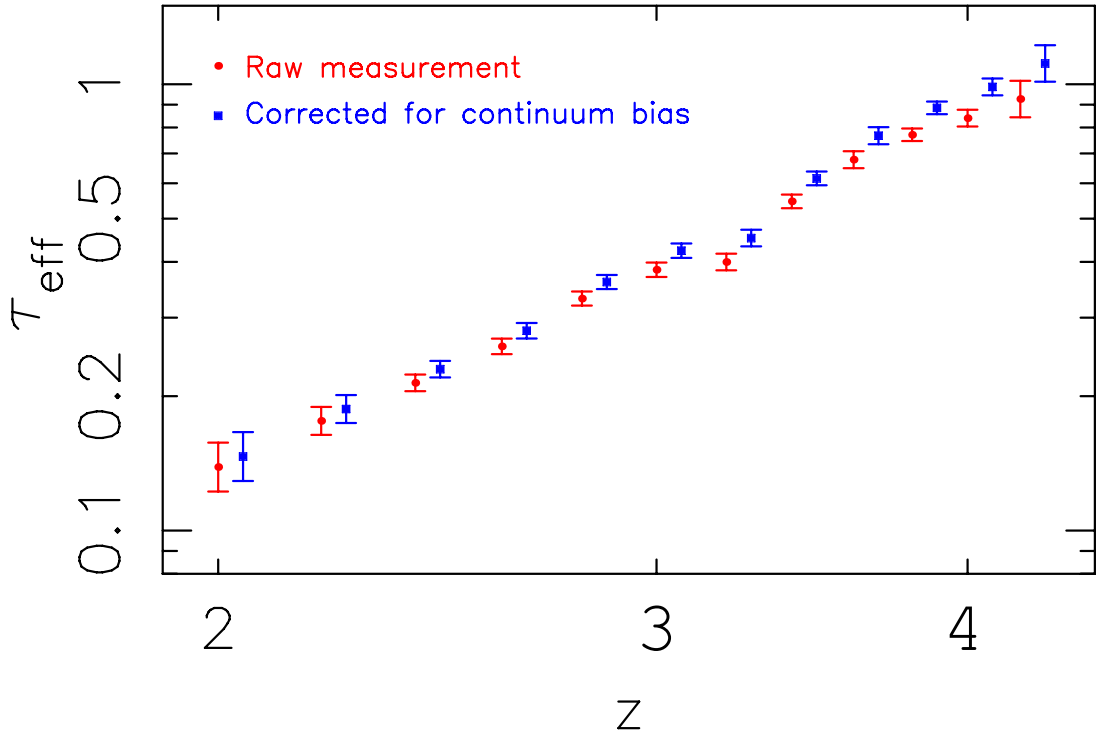}
\end{center}
\caption{Measurement of $\tau_{\rm eff}$ vs. $z$ in $\Delta z=0.2$ bins from the complete data set.
The red circles show the raw measurement obtained from the continuum fits.
The blue squares show the measurement after correction for continuum bias as in \S \ref{continuum estimation}.
Error bars are one standard deviation and calculated as in \S \ref{error bars}.
The continuum-corrected data points are centered on the same redshifts as the raw measurement, but have been slightly offset to the right for graphical clarity.
Numerical values are provided in Table \ref{all measurements table}.}
\label{cont corr comparison}
\end{figure*}

The continuum level of each spectrum is estimated by fitting a cubic spline through its transmission peaks for the ESI and MIKE spectra.
To do so, we use the IDL graphical user interface {\tt x\_continuum} developed by J.~X. Prochaska.
More precisely, we manually identify peaks of transmission near unity by looking for plateaux consisting of several consecutive pixels with approximately the same transmission level and force the spline to pass through these points.
In practice, these plateaux have a finite thickness owing to the noise level and so we place the points at the centers of the noise bands.
Additional spline points are then placed and adjusted to approximate the most featureless continuum consistent with the identified transmission peaks. 
In general, the number of transmission peaks that we are able to identify decreases with redshift.
The robustness and accuracy of this manual spectrum-by-spectrum continuum estimation method are tested in \S \ref{continuum estimation} and Appendix \ref{continuum fitting appendix}.
For the HIRES data, the continua were similarly estimated as part of the procedure to join the different echelle orders. 
Figures \ref{example spectra} and \ref{example cont fits} show examples of continuum fits on real and mock spectra, respectively.
The masks, detailed in the next section, are necessary to ensure that the measurement is uncontaminated.
Pixels, which actually have a finite width, are labeled by their central redshift in performing the binning.
Any given pixel therefore falls in the bin which contains $>50$\% of it.
Errors are calculated as in \S \ref{error bars}.
Our measurement from the complete data set (HIRES+ESI+MIKE) in $\Delta z=0.2$ redshift bins, raw and corrected for continuum bias as in \S \ref{continuum estimation}, is shown in Figure \ref{cont corr comparison}.
The numerical values are provided in Table \ref{all measurements table}, in which we compile all of our $\tau_{\rm eff}$ measurements, corrected for metal absorption (see \S \ref{metal absorption}) and not, both in $\Delta z=0.1$ and $\Delta z=0.2$ redshift bins.

\begin{deluxetable*}{ccccccccc}
\tablewidth{0pc}
\tablecaption{Measurement of $\tau_{\rm eff}$, with and without
continuum and metal corrections, and in redshift bins of width $\Delta z=0.2$ and $\Delta z=0.1$\label{all measurements table}}
\tabletypesize{\footnotesize}
\tablehead{
\colhead{z} &
\colhead{Raw\tablenotemark{a}} &
\colhead{Cont. corr.\tablenotemark{b}} &
\colhead{Schaye metal corr.\tablenotemark{c}} &
\colhead{Kirkman metal corr.\tablenotemark{c}} &
\colhead{$\sigma_{\tau_{\rm eff},\rm stat}$\tablenotemark{d}} & 
\colhead{$\sigma_{\tau_{\rm eff},\rm cont}$\tablenotemark{e}} &
\colhead{$\sigma_{\tau_{\rm eff},\rm metal}$\tablenotemark{f}} &
\colhead{$\sigma_{\tau_{\rm eff},\rm tot}$\tablenotemark{g}} \\
\hline \multicolumn{9}{c}{$\Delta z=0.2$}
}
\startdata
2.0 & 0.138 & 0.146 & 0.127 & 0.115 & 0.018 & 0.002 & 0.012 & 0.023 \\
2.2 & 0.176 & 0.187 & 0.164 & 0.157 & 0.013 & 0.003 & 0.014 & 0.020 \\
2.4 & 0.214 & 0.229 & 0.203 & 0.200 & 0.009 & 0.004 & 0.016 & 0.019 \\
2.6 & 0.258 & 0.280 & 0.251 & 0.251 & 0.010 & 0.006 & 0.018 & 0.022 \\
2.8 & 0.330 & 0.360 & 0.325 & 0.332 & 0.012 & 0.008 & 0.021 & 0.026 \\
3.0 & 0.384 & 0.423 & 0.386 & 0.396 & 0.014 & 0.010 & 0.022 & 0.029 \\
3.2 & 0.399 & 0.452 & 0.415 & 0.425 & 0.017 & 0.012 & 0.022 & 0.031 \\
3.4 & 0.546 & 0.615 & 0.570 & 0.589 & 0.019 & 0.017 & 0.027 & 0.039 \\
3.6 & 0.677 & 0.766 & 0.716 & 0.742 & 0.030 & 0.023 & 0.030 & 0.051 \\
3.8 & 0.770 & 0.885 & 0.832 & 0.861 & 0.025 & 0.028 & 0.031 & 0.051 \\
4.0 & 0.839 & 0.986 & 0.934 & 0.963 & 0.037 & 0.034 & 0.031 & 0.064 \\
4.2 & 0.926 & 1.113 & 1.061 & 1.090 & 0.091 & 0.041 & 0.031 & 0.121 \\
\hline \multicolumn{9}{c}{$\Delta z=0.1$} \\ \hline
2.0 & 0.148 & 0.156 & 0.135 & 0.125 & 0.022 & 0.002 & 0.012 & 0.027 \\
2.1 & 0.130 & 0.139 & 0.121 & 0.109 & 0.017 & 0.002 & 0.011 & 0.022 \\
2.2 & 0.173 & 0.184 & 0.161 & 0.154 & 0.016 & 0.003 & 0.013 & 0.023 \\
2.3 & 0.230 & 0.243 & 0.215 & 0.213 & 0.018 & 0.004 & 0.017 & 0.026 \\
2.4 & 0.184 & 0.199 & 0.177 & 0.170 & 0.010 & 0.004 & 0.014 & 0.018 \\
2.5 & 0.238 & 0.257 & 0.229 & 0.228 & 0.012 & 0.005 & 0.017 & 0.022 \\
2.6 & 0.269 & 0.291 & 0.260 & 0.262 & 0.013 & 0.006 & 0.018 & 0.024 \\
2.7 & 0.284 & 0.309 & 0.278 & 0.281 & 0.015 & 0.006 & 0.019 & 0.026 \\
2.8 & 0.334 & 0.363 & 0.328 & 0.335 & 0.018 & 0.008 & 0.021 & 0.030 \\
2.9 & 0.339 & 0.373 & 0.339 & 0.346 & 0.016 & 0.008 & 0.021 & 0.029 \\
3.0 & 0.401 & 0.441 & 0.402 & 0.413 & 0.021 & 0.010 & 0.023 & 0.035 \\
3.1 & 0.405 & 0.450 & 0.412 & 0.424 & 0.023 & 0.011 & 0.023 & 0.036 \\
3.2 & 0.402 & 0.454 & 0.418 & 0.428 & 0.023 & 0.012 & 0.022 & 0.036 \\
3.3 & 0.419 & 0.479 & 0.442 & 0.453 & 0.025 & 0.013 & 0.022 & 0.038 \\
3.4 & 0.533 & 0.602 & 0.558 & 0.576 & 0.025 & 0.017 & 0.026 & 0.043 \\
3.5 & 0.685 & 0.763 & 0.710 & 0.738 & 0.029 & 0.022 & 0.032 & 0.051 \\
3.6 & 0.666 & 0.755 & 0.705 & 0.731 & 0.046 & 0.023 & 0.030 & 0.065 \\
3.7 & 0.718 & 0.820 & 0.768 & 0.795 & 0.042 & 0.025 & 0.030 & 0.062 \\
3.8 & 0.753 & 0.868 & 0.817 & 0.844 & 0.033 & 0.028 & 0.031 & 0.057 \\
3.9 & 0.802 & 0.931 & 0.879 & 0.908 & 0.029 & 0.031 & 0.031 & 0.056 \\
4.0 & 0.782 & 0.929 & 0.880 & 0.906 & 0.058 & 0.032 & 0.029 & 0.082 \\
4.1 & 0.953 & 1.118 & 1.063 & 1.095 & 0.082 & 0.040 & 0.033 & 0.110 \\
4.2 & 0.893 & 1.080 & 1.030 & 1.058 & 0.101 & 0.040 & 0.030 & 0.132 \\
\enddata
\tablenotetext{a}{Estimated directly from the continuum fits as in \S \ref{definitions}.}
\tablenotetext{b}{Raw measurement corrected for the redshift-dependent bias in the continuum fits as in \S \ref{continuum estimation}.}
\tablenotetext{c}{Same as the measurement corrected for continuum bias, but with the contribution of metals to absorption in the \Lya~forest corrected as in \S \ref{metal absorption}.} 
\tablenotetext{d}{Statistical error, estimated as in \S \ref{error bars}.}
\tablenotetext{e}{Systematic error arising from the continuum correction, estimated as in \S \ref{continuum bias systematic error}.}
\tablenotetext{f}{Systematic error arising from the metal correction, estimated as in \S \ref{metal correction systematic error}.}
\tablenotetext{g}{Quadrature sum of the statistical and systematic errors (\S \ref{total error}).}
\end{deluxetable*}

\subsection{Masks}
\label{masks}
\subsubsection{Definition of the \Lya~Forest}
\label{forest definition}
To avoid contamination from absorption by higher Lyman-series resonances, we consider only those pixels between the rest-frame wavelengths of Ly$\beta$ (1025 \AA) and Ly$\alpha$ (1216 \AA).
Further, we mask the ``proximity region'' of each quasar, in which the absorption is biased low with respect to the cosmic mean owing to the ionizing radiation of the quasar itself.
Specifically, we mask the 25 proper Mpc nearest to each quasar.
The HI-ionizing radiation of bright quasars typically equals the cosmic mean at distances $\sim5$ proper Mpc \cite[e.g.,][]{2007astro.ph..1042F}, so that our masks are very conservative and little contamination should remain.

\subsubsection{Bad and Low-$S/N$ Pixels}
Some quasar spectra contain bad pixels, resulting in gaps in the data.
These bad pixels are identified by carefully inspecting each spectrum and are then masked.
The sensitivity of the spectrographs generally degrades at shorter wavelengths.
The correspondingly low $S/N$ makes reliable continuum estimation challenging.
We thus keep only the portions of the spectra where we judge that the continuum can be reliably estimated, corresponding to $S/N\gtrsim10$ per pixel.

\subsubsection{Damped Absorbers}
Most of the quasar spectra used here were obtained as part of studies of SLLS and DLAs.
Since these systems individually contribute large equivalent widths to the observed absorption, we mask these systems to ensure that our measurement of the \Lya~forest mean transmission is unbiased.
Specifically, we mask systems with $\log_{10} N_{\rm HI}\geqslant19.0$ and associated metal lines.
We show in Appendix \ref{dla tests} that these systems contribute only about 2\% of the total absorption in the \Lya~forest in an unbiased sample. 
This is the amount by which the absorption is underestimated by masking these systems in our biased sample.  
The width of each mask is chosen such that all data around the SLLS or DLAs where the transmission in the presence of that system only would be $<99\%$ is ignored.
Beyond masking the absorption from the SLLS or DLAs, this helps to keep the error on the estimated continuum owing to the damping wings $\lesssim1\%$.
In Appendix \ref{dla tests}, we increase the width of the masks on the SLLS and DLAs by a factor of two and conclude that our measurement is robust to the choice of width.
For the metal lines, we mask the most common neutral species (MgI, OI, etc.), low-ionization ions -- which are dominant in HI absorbers - (FeII, SiII, CII, OI, NI, AlII, etc.), intermediate ions (AlIII, FeIII, CIII, SiIII, etc.), and high-ionization ions (SiIV, CIV, NV, OVI, etc.).
We place a mask of width 500 km s$^{-1}$ centered on each metal line associated with a SLLS or DLA in our sample.
The width of metal lines is generally dominated by Doppler broadening owing to peculiar motion, for which the median is $\sim$80 km s$^{-1}$ \citep{1997ApJ...474..140P}.
Our masks are thus sufficiently large to completely masks these lines and allow for an uncertainty $\sim200$ km s$^{-1}$ in each direction on the redshift of the lines.
We show in Appendix \ref{dla tests} that masking metal lines associated with DLAs has practically no effect on our measurement.

\subsection{Statistical Error Bars}
\label{error bars}
To estimate the statistical uncertainty on our measurement of $\langle F \rangle$, we proceed as follows.
For any given redshift bin, we first concatenate the data from the different spectra.
We then break the total data stream in segments of length 3 proper Mpc, where the length is chosen such that the mean transmission calculated from the i-th segment, $\langle F \rangle_{i}$, is independent of its neighbors.
\cite{2000ApJ...543....1M} and \cite{2006ApJS..163...80M} measured the flux correlation function and power spectrum in the \Lya~forest and found small correlations on scales $\gtrsim$1 proper Mpc.
\cite{2007astro.ph..1042F} computed the optical depth correlation coefficient $\rho$ from a simulation of the \Lya~forest and found $\rho<1\%$ for $r>3$ proper Mpc over the redshift range $2\leq z \leq4$.
The error on $\langle F \rangle$ is then estimated as 
\begin{equation}
\sigma_{\langle F \rangle}=
\frac{\sigma_{\langle F \rangle_{i}}}{\sqrt{N}},
\end{equation}
where $\sigma_{\langle F \rangle_{i}}$ is the standard deviation among the individual $\langle F \rangle_{i}$, and $N$ is the number of independent segments.
This estimate simply makes use of the fact that $\langle F \rangle$ is the mean of the $\langle F \rangle_{i}$, which are mutually independent, and of the usual result that the error on the mean of independent measurements is just equal to their standard deviation divided by $\sqrt{N}$.
We have verified that the error estimates have converged for our choice of segment length.
\begin{figure*}[ht]
\begin{center}
\includegraphics[width=1.00\textwidth]{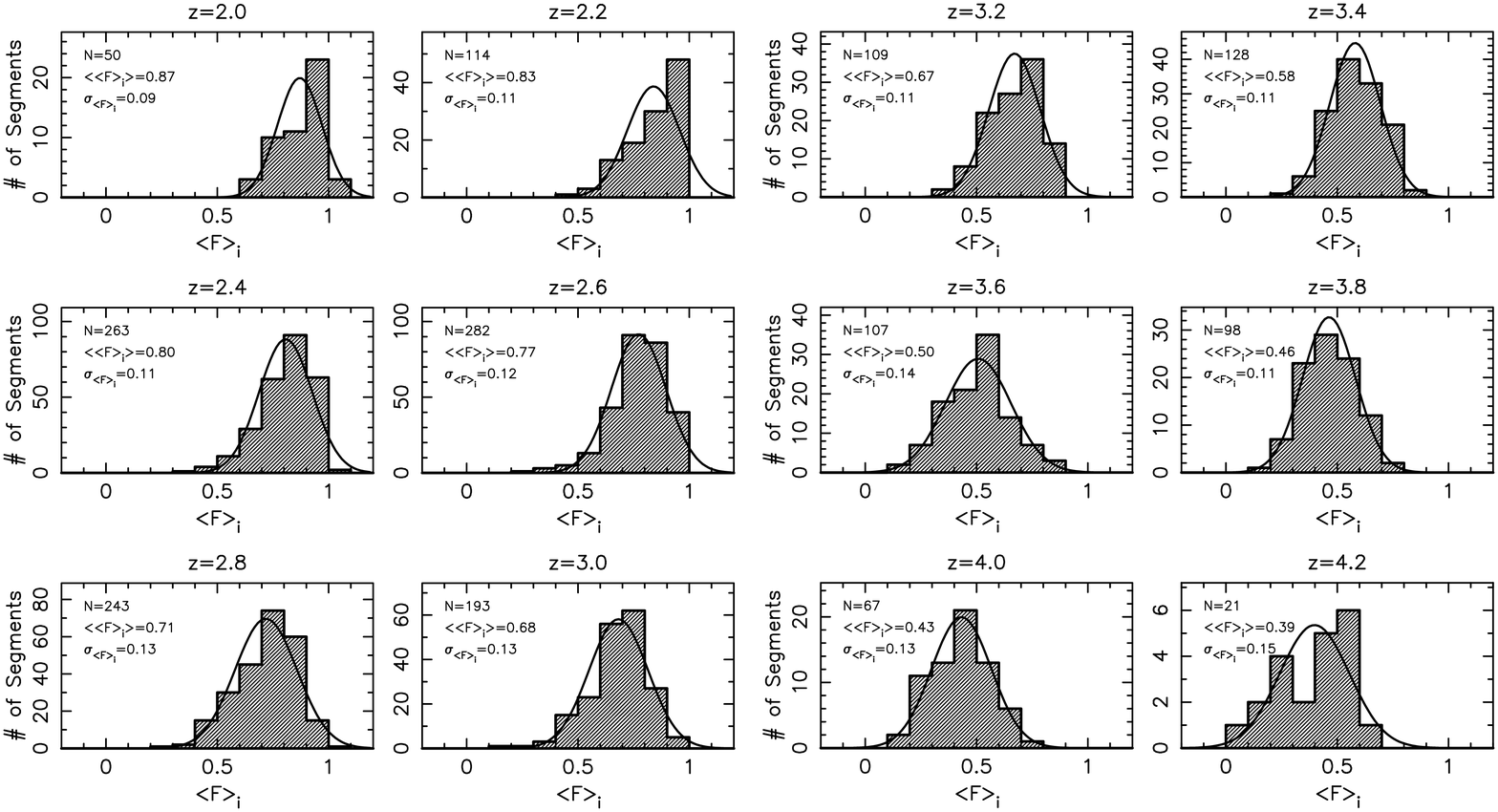}
\end{center}
\caption{Histograms of the segment-averaged transmissions $\langle F \rangle_{i}$.
Each panel corresponds to a redshift bin of width $\Delta z=0.2$.
The text in the upper left corners indicates the total number of 3 proper Mpc segments in the bin, the mean $\langle F \rangle_{i}$, and their standard deviation.
The 1$\sigma$ uncertainty on $\langle F \rangle=\langle \langle F \rangle_{i} \rangle$ is estimated as $\sigma_{\langle F \rangle}=\sigma_{\langle F \rangle_{i}}/\sqrt{N}$ (\S \ref{error bars}).
The solid curves shows the Gaussian function with same mean and standard deviation as the histograms.
\label{chunk hists}}
\end{figure*}
In the limit of an infinite number of segments, the central limit theorem ensures that the uncertainty on $\langle F \rangle$ is Gaussian.
Figure \ref{chunk hists} shows histograms of $\langle F \rangle_{i}$ for the redshift bins considered.
The histograms are close to Gaussian, so that the convergence to Gaussianity of the error on $\langle F \rangle$ should be rapid, and may thus confidently be expected to have been attained.
The histograms of Figure \ref{chunk hists} also show no sign of outliers, arguing that our measurement is unaffected by them.\\ \\
In \S \ref{systematic errors}, we estimate the systematic error budget, including contributions from the continuum and metal corrections (see below).

\subsection{Continuum Bias}
\label{continuum estimation}
Our measurement of $\langle F \rangle$ is a direct function of our continuum estimates, and we must therefore ask how accurate these are.

Random continuum errors are included in the statistical error estimates described \S \ref{error bars}.
We thus need only further consider potential systematic biases: finite resolution and $S/N$ of the data, and a possible redshift-dependent bias arising from increasing \Lya~absorption with increasing redshift.

\begin{figure}[ht]
\begin{center}
\includegraphics[width=0.40\textwidth]{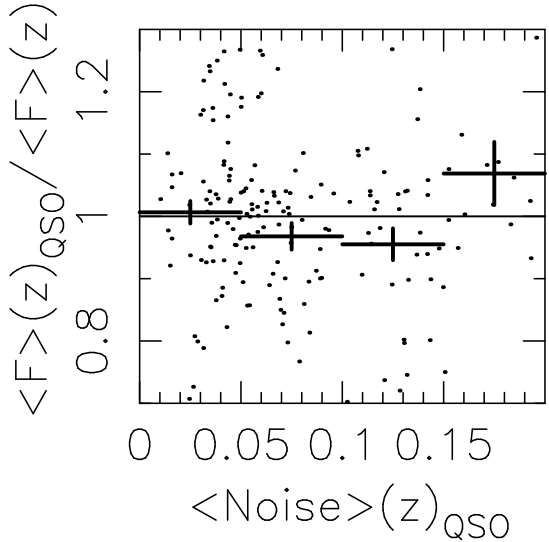}
\end{center}
\caption{Ratios of $\langle F \rangle(z)$ as estimated from every particular quasar intersecting a given redshift bin, $\langle F \rangle(z)_{QSO}$, to $\langle F \rangle(z)$ for the entire bin (as calculated from all the quasars contributing to it) versus the mean noise on $F$ in the particular quasar's data in the bin.
The crosses show averages over noise bins of width 0.05.
The 1$\sigma$ vertical error bars are equal the standard deviation of the points in the bin divided by the square root of the number of points.
No significant trend with noise is seen, arguing against noise-dependent systematic effects.
}
\label{noise scatter}
\end{figure}

Finite resolution and $S/N$ both make accurate continuum fitting more challenging.
In the case of finite $S/N$, the pixel-wise ``peaks'' in the observed spectra in general do not exactly correspond to a transmission of unity, owing to the noise contribution.
Since the noise fluctuations are symmetric about zero and we attempt to identify the transmission peaks with the centers of noise bands, finite $S/N$ is not expected to contribute a significant systematic bias, provided $S/N\gtrsim10$ per pixel.
To verify this, we plot the ratios of $\langle F \rangle(z)$ as estimated from every particular quasar intersecting a given redshift bin, $\langle F \rangle(z)_{QSO}$, to $\langle F \rangle(z)$ for the entire bin (as calculated from all the quasars contributing to it) versus the mean noise in the particular quasar's data in the bin in Figure \ref{noise scatter}.
We use redshift bins of width $\Delta z=0.2$.
These ratios should average to unity in the absence of systematics.
No significant trend with noise is seen, thus arguing against noise-dependent systematic effects.

In the case of finite resolution, the peaks are smoothed and their height may therefore be underestimated.
Is this effect important for our data?
The pressure of the baryons smoothes their spatial distribution on a scale of order the Jeans scale,
\begin{equation}
\lambda_{J} =
\sqrt{
\frac{\pi \gamma k T}{G \rho \mu m_{p}},
}
\end{equation}
where $\gamma$ is the adiabatic index of the gas, $T$ is its temperature, $\rho$ is the total mass density, $\mu$ the mean molecular weight of the gas, and $m_{p}$ is the proton mass.
To convert to km s$^{-1}$, we multiply $\lambda_{J}$ by $H(z)=\sqrt{8\pi G \rho/3}$ (the Friedmann equation for a flat universe).
For $\gamma=1.62$ \citep[valid in the limit of early reionization;][]{1997MNRAS.292...27H}, $\mu=0.59$ (for a fully ionized gas consisting of 75\% hydrogen and 25\% helium by mass) and $T=20,000$ K \citep{2000ApJ...534...41R, 2000MNRAS.318..817S, 2001ApJ...562...52M, 2001ApJ...557..519Z}, $\lambda_{J}\approx103$ km s$^{-1}$, larger than the resolution of each spectrograph used in this study (c.f. \S \ref{data set}).
If the transmission peaks correspond to density troughs of size at least the Jeans scale, this suggests that all of our spectra will resolve them, and hence that our measurement will not be affected by any significant resolution-dependent bias.
This argument, however, has caveats.

The correct baryonic smoothing scale is in fact not equal to the Jeans scale, but instead is expected to be smaller, owing to the finite time necessary for the baryons to respond to the heat injected into the IGM during HI reionization at redshift $\gtrsim6$ \citep{1998MNRAS.296...44G}.
On the other hand, thermal broadening redshift-space distortions will additionally smooth the spectra on a scale comparable to the Jeans scale \citep[for details about redshift-space distortions in the \Lya~forest, see e.g.][]{2007astro.ph..1042F}. 
The \Lya~forest transmitted flux is also measured to have non-zero power on scales smaller than the Jeans \citep{2000ApJ...543....1M}.
Further, the IGM temperature in underdense regions is generally below the cosmic mean, reducing the relevant local smoothing scale, and may not be definitively known.
\begin{figure}[ht]
\begin{center}
\includegraphics[width=0.40\textwidth]{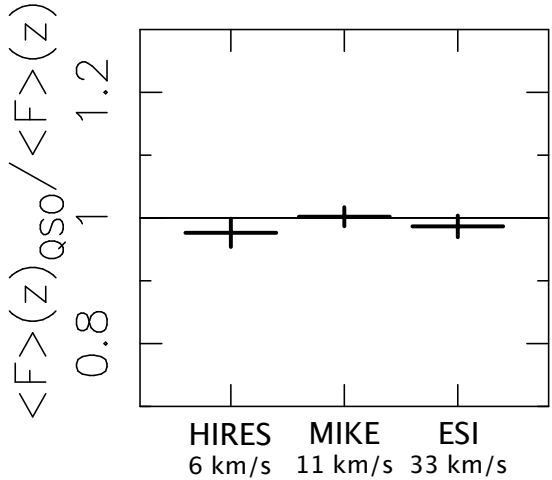}
\end{center}
\caption{Mean ratio of $\langle F \rangle(z)$ as estimated from every particular quasar intersecting a given redshift bin, $\langle F \rangle(z)_{QSO}$, to $\langle F \rangle(z)$ for the entire bin (as calculated from all the quasars contributing to it) versus the instrument with which each spectrum was obtained.
The $FWHM$ of each spectrograph is indicated on the figure (\S \ref{data set}).
The true resolution varies from spectrum to spectrum, depending on the width of the slit used; here we show an optimistic, yet representative, value for each spectrograph.
The 1$\sigma$ vertical error bars are equal the standard deviation of the points in the bin divided by the square root of the number of points.
No significant instrument-specific bias or trend with resolution is seen, arguing against such systematic effects.
}
\label{spectro scatter}
\end{figure}
To test for resolution- or spectrograph-dependent systematics from the data itself, we compare in Figure \ref{spectro scatter} the mean ratio of $\langle F \rangle(z)$ as estimated from every particular quasar intersecting a given redshift bin, $\langle F \rangle(z)_{QSO}$, to $\langle F \rangle(z)$ for the entire bin (as calculated from all the quasars contributing to it) versus the instrument with which each spectrum was obtained.
Again, no significant bias or trend with instrument or resolution is seen.

\begin{figure}[ht]
\begin{center}
\includegraphics[width=0.45\textwidth]{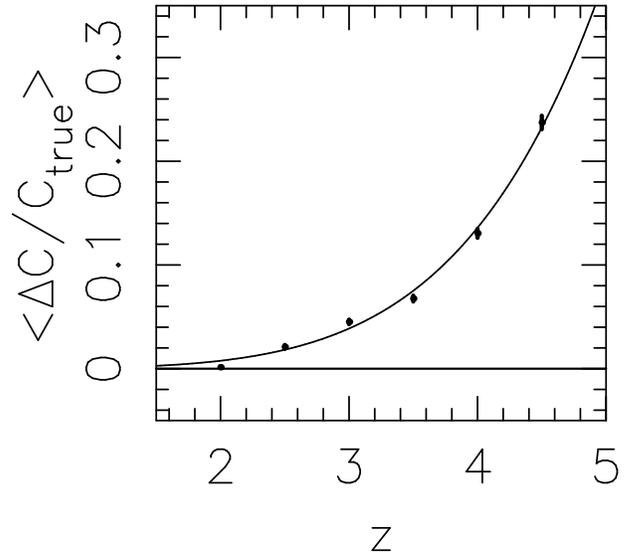}
\end{center}
\caption{Mean fractional difference between the estimated and true continua as a function of redshift.
The points at redshift intervals $\Delta z=0.5$ were calculated by comparing blind continuum fits on mock spectra (see Figure \ref{example cont fits}) to true continua for 10 mock spectra at each redshift, as described in \S \ref{continuum estimation} and Appendix \ref{continuum fitting appendix}.
The small error bars indicate the 1$\sigma$ errors on the means.
The redshift trend is well approximated by the best-fit power law $\Delta C/C_{\rm true}=1.58\times10^{-5}(1+z)^{5.63}$ over the redshift range covered.
}
\label{continuum trend}
\end{figure}

Finally, as the redshift increases, the IGM density increases owing to the cosmological expansion.
As a result, the \Lya~absorption also increases with redshift and transmission peaks reaching unity become increasingly rare, potentially causing us to underestimate the continuum level more at higher redshifts. 
In Appendix \ref{continuum fitting appendix}, we detail how we use mock spectra to quantify the mean fractional difference between the true and estimated continua, $\langle \Delta C/C_{\rm true} \rangle$ (where $\Delta C = C_{\rm est}-C_{\rm true}$), as function of redshift.
We find that, for hydrogen background photoionization rates (\Gbkg) taken from the literature, $\Delta C/C_{\rm true}=1.58\times10^{-5}(1+z)^{5.63}$ to sub-percent statistical precision over the redshift range $2\leq z \leq4.5$, as illustrated in Figure \ref{continuum trend}.
We note, however, that this continuum correction itself is subject to some systematic bias, since it depends on assumptions (particularly regarding \Gbkg) used to generate the mock spectra.
This potential systematic bias is explored in detail in the Appendix. 
In the Appendix, we also show that if $\tau_{\rm eff}^{\rm est}$ is the effective optical depth as estimated as above, and $\tau_{\rm eff}^{\rm true}$ is the ``true'' value after correcting for the redshift-dependent continuum bias, then
\begin{equation}
\tau_{\rm eff}^{\rm true} = \tau_{\rm eff}^{\rm est} - 
\ln{\left[
1 - 
\left\langle
\frac{\Delta C}{C_{\rm true}}  
\right\rangle
\right]}.
\end{equation}
We use this expression to correct our measurement for continuum error; the continuum-corrected measurement is compared to the raw measurement in Figure \ref{cont corr comparison}. 
We will henceforth generally consider this continuum-corrected measurement.\\ \\
Of course, this effect does not affect measurements based on extrapolating the continuum from redward of the \Lya~line, as usually done in measurements based on lower resolution data for which direct local continuum estimation is impossible \citep[e.g.,][]{1993ApJ...414...64P, 2003AJ....125...32B, 2005ApJ...635..761M}.
Such extrapolations, however, likely miss local variations in the continuum from quasar to quasar so that local estimation is advantageous for high-resolution data.

\subsection{Metal Absorption}
\label{metal absorption}
Metals also contribute to the total absorption in the \Lya~forest.
Consider the transmission at observed wavelength $\lambda$, $F(\lambda)$, in a given spectrum.
Generically labeling metals by $m$,
\begin{equation}
F(\lambda) \equiv 
F_{\alpha} \times 
\prod_{m} F_{m}
\equiv
e^{-\tau_{\alpha}} \times
\prod_{m} e^{-\tau_{m}},
\end{equation}
where $\tau_{\alpha}$ is the \Lya~optical depth and $\tau_{m}$ is the optical depth owing to metal $m$.\footnote{For the purpose of this discussion, different metal ionization states are viewed as different species.}
Because each metal transition has a different rest-frame wavelength, absorption manifesting itself at observed wavelength $\lambda$ occurs in physically separated regions of the IGM.
The $F_{m}$ are therefore statistically independent, so
\begin{equation}
\langle F(\lambda) \rangle \equiv 
\langle F_{\alpha} \rangle
\times 
\prod_{m} 
\langle F_{m} \rangle.
\end{equation}
From this,
\begin{equation}
\label{tau eff tot}
\tau_{\rm eff} = \tau_{{\rm eff},\alpha} + \sum_{m} \tau_{{\rm eff},m},
\end{equation}
where we have simply generalized the effective optical depth definition (eq. \ref{tau eff}).
Equation (\ref{tau eff tot}) states that the measured effective optical depth is actually the sum of a contribution from pure \Lya~and contributions from metals.\\ \\
The easiest quantity to model, either analytically or by simulation, is $\tau_{{\rm eff},\alpha}$.

Can we correct our measurement to estimate $\tau_{{\rm eff},\alpha}$? 
To a certain extent, it is possible to separate metal absorption from \Lya~absorption.
We explore two ways, one direct and one statistical, based on published measurements.

\cite{2003ApJ...596..768S} measured $\tau_{\rm eff}$ from 19 high-quality spectra obtained with the UV-Visual Echelle Spectrograph (UVES) instrument on the Very Large Telescope (VLT) and HIRES over approximately the same redshift range as us ($2\lesssim z\lesssim 4.2$).
These authors attempted to directly identify contaminating metal lines by searching by eye for absorption features corresponding to the redshifts of strong HI lines and to the redshifts of the absorption systems redward of the quasar's \Lya~emission line.
They fitted a power law of the form
\begin{equation}
\log{\tau_{\rm eff}} = \log{\tau_{0}} + \alpha \log{[(1+z)/4]}
\end{equation}
to both the data with all the metals and with metals removed. 
When metals are kept, they found $(\log{\tau_{0}}, \alpha) = (-0.40 \pm 0.02, 3.40 \pm 0.23)$, and when metals are masked $(\log{\tau_{0}}, \alpha) = (-0.44 \pm 0.01, 3.57 \pm 0.20)$.
We must caution in passing that the 1$\sigma$ error bars on the \cite{2003ApJ...596..768S} $\tau_{\rm eff}$ measurements appear to be significantly underestimated based on comparing the scatter between the different data points to their reported errors (see their Figure 1).
Assuming that the relation between the \cite{2003ApJ...596..768S} fits accurately represents the metal contribution to the absorption in the \Lya~forest, we can calculate a correction to our more precise $\tau_{\rm eff}$ measurement:
\begin{equation}
\label{schaye metal correction}
\tau_{{\rm eff},\alpha} = 
\frac{\tau_{0,{\rm no~metals}}}{\tau_{0,{\rm metals}}}
\times
\left(
\frac{1+z}{4}
\right)^{(\alpha_{{\rm no~metals}} - \alpha_{{\rm metals}})} 
\times
\tau_{{\rm eff}}.
\end{equation}

\begin{figure*}[ht]
\begin{center}
\includegraphics[width=0.80\textwidth]{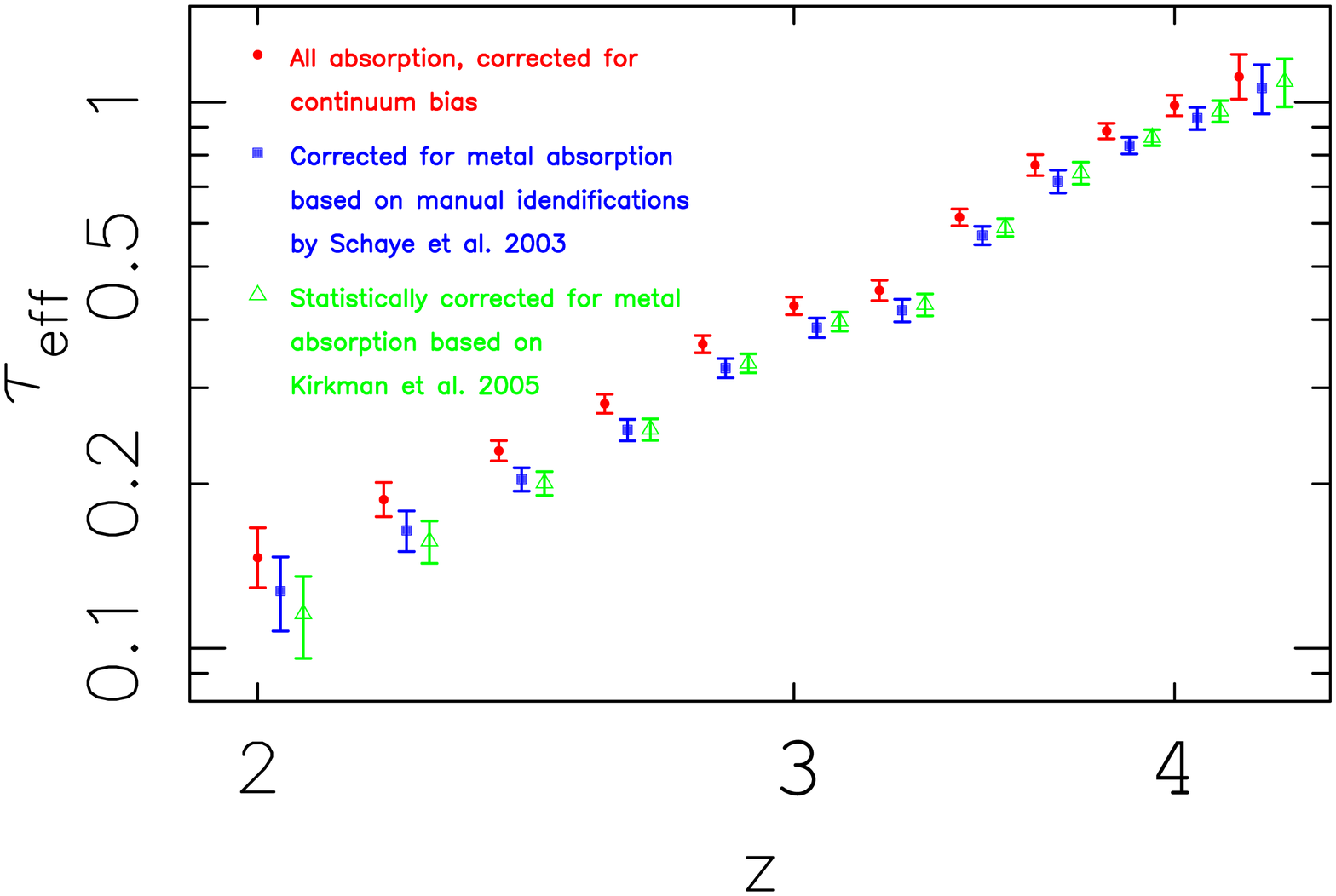}
\end{center}
\caption{Comparison of our measurement of the total $\tau_{\rm eff}$ (red circles) with the inferred values including \Lya~absorption only, $\tau_{{\rm eff},\alpha}$, with the metal contribution estimated based on the fits of \cite{2003ApJ...596..768S} (blue squares), who manually masked metal lines in their spectra (left), and with the metal contribution statistically estimated based on the fits of \cite{2005MNRAS.360.1373K} (green triangles).
The metal-corrected data points have been slightly offset to the right for graphic clarity, but they are actually centered on the same redshifts as the uncorrected $\tau_{\rm eff}$ measurement.
All measurements have been corrected for continuum bias, as described in \S \ref{continuum estimation}.
Numerical values are provided in Table \ref{all measurements table}.
}
\label{metals vs no metals figure}
\end{figure*}

In Figure \ref{metals vs no metals figure}, we compare our $\tau_{\rm eff}$ measurement with $\tau_{{\rm eff},\alpha}$ inferred using this correction.
The relative metal correction to $\tau_{\rm eff}$ decreases with increasing redshift, as expected if the IGM is monotonically enriched with time\footnote{Note, however, that the metal contribution to the absorption (i.e., $\tau_{\rm eff}$) need not strictly follow the abundance of the metals, since the amount of absorption also depends on the ionization state of the metals, which in turn depends on parameters like the intensity and hardness of the radiation background.}: 13\% at $z=2$, 9\% at $z=3$, and 5\% at $z=4$. 

An alternative to directly identifying metal lines in the \Lya~forest is to statistically remove them using measurements of ``pure'' metal absorption redward of the \Lya~emission line.
\cite{2004AJ....128.1058T} applied this idea in two ways at redshift $\sim1.9$.
First, they measured the absorption in the rest-frame wavelength range 1225-1500 \AA~in their sample of spectra obtained with the Kast spectrograph on the Lick Observatory 3 m telescope.
As more metal lines contribute to the absorption at shorter rest-frame wavelengths, they extrapolated to estimate the metal absorption in the \Lya~forest wavelength range 1070-1170 \AA.
As a check, \cite{2004AJ....128.1058T} also summed the equivalent widths of metal absorption lines for 26 quasars with $1.7<z_{em}<2.3$ tabulated by \cite{1988ApJS...68..539S} to estimate the amount of absorption by metals.
\cite{2004AJ....128.1058T} considered the quantity $DA\equiv1-\langle F \rangle$ instead of $\tau_{\rm eff}$, but note that $\langle F \rangle \equiv \exp{(-\tau_{\rm eff})}\approx1 - \tau_{\rm eff}$ and hence $\tau_{\rm eff}\approx DA$ for $\tau_{\rm eff} \ll 1$, which is a good approximation at $z\sim2$.
Defining $DM$ to be the contribution to $DA$ owing to metals only, \cite{2004AJ....128.1058T} found $DM=0.158\pm0.13$ from their Kast analysis (their $DM2$) and $0.165\pm0.22$ (their $DM6$) using the lines tabulated by \cite{1988ApJS...68..539S} at observed wavelength 4158 \AA.
The two values thus agree very well.
Building upon the work of \cite{2004AJ....128.1058T}, \cite{2005MNRAS.360.1373K} included all 52 \cite{1988ApJS...68..539S} quasars to estimate the metal absorption over the extended redshift range $1.7<z<3.54$. 
They estimated the metal contribution as a function of both rest-frame ($\lambda_{r}$) and observed wavelengths ($\lambda_{o}$).
In their notation:
\begin{equation}
\label{dm5}
DM5(\lambda_{r})=0.01564 - 4.646\times10^{-5}(\lambda_{r}-1360~\textrm{\AA})
\end{equation}
and
\begin{equation}
\label{dm6}
DM6(\lambda_{o})=0.01686 - 1.798\times10^{-6}(\lambda_{o}-4158~\textrm{\AA}).
\end{equation}
Although the observed wavelength is the relevant quantity for the redshift of \Lya~absorption, the trend with rest-frame wavelength serves to approximately account for the growing collection of metals contributing absorption at shorter wavelengths.
We first estimate the metal contribution at the mean rest-frame wavelength of the \cite{1988ApJS...68..539S} sample, $\langle \lambda_{r} \rangle=1360$~\AA, and at \Lya~redshift $z$ using Equation (\ref{dm6}):
\begin{align}
\label{kirkman correction redshift}
\tau_{{\rm eff},m}(z=\lambda_{o}/1216\textrm{ \AA}-1;~\lambda_{r}=1360\textrm{ \AA}) &= \notag \\ 
-\ln{[1-DM6(\lambda_{o})]}.
\end{align}
We then use Equation (\ref{dm5}) to extrapolate $\tau_{{\rm eff},m}$ to a wavelength approximately in the center of the \Lya~forest region, 1120 \AA:
\begin{align}
\label{kirkman correction wavelength}
\frac{\tau_{{\rm eff},m}(z;~\lambda_{r}=1120~\textrm{\AA})}
{\tau_{{\rm eff},m}(z;~\lambda_{r}=1360~\textrm{\AA})} &=
\frac{\ln{[1-DM5(\lambda_{r}=1120~\textrm{\AA})]}}
{\ln{[1-DM5(\lambda_{r}=1360~\textrm{\AA})]}} \notag \\
&=1.723.
\end{align}

In Figure \ref{metals vs no metals figure}, we compare our $\tau_{\rm eff}$ measurement corrected only for continuum bias with both $\tau_{{\rm eff},\alpha}$ inferred using this statistical method correction and with the direct method based on the \cite{2003ApJ...596..768S} fits.
The fraction of $\tau_{\rm eff}$ contributed by metals is $21\%$ at $z=2$, $6\%$ at $z=3$, and $2\%$ at $z=4$ according to the statistical metal correction.
Our metal-corrected $\tau_{{\rm eff},\alpha}$ estimates based on removing metals directly and statistically agree well within 1$\sigma$ at each redshift considered, thus suggesting that the metal correction using either method is accurate to the level of our statistical error bars, or better.
A more detailed estimate of the uncertainty on $\tau_{\rm eff,\alpha}$ introduced by the metal correction is provided in \S \ref{systematic errors}. \\ \\
Numerical values are tabulated in Table \ref{all measurements table}.\\ \\

\section{SYSTEMATIC ERRORS}
\label{systematic errors}
In the Appendix, we present tests arguing that our measurement is free of several systematic effects, including effects arising from the data being obtained with different instruments; Ly$\beta$, OVI, and associated absorption; and SLLS and DLAs.
We also discuss how our continuum correction may itself suffer from some level of bias.
However, our error bars thus far have not accounted for the systematic biases that may affect both the continuum and metal corrections.
These biases are difficult to accurately quantify.
Moreover, they are most likely smoothly (thus affecting neighboring bins in with strong correlation) and monotonically varying with redshift.
For the purpose of investigating whether the redshift evolution of $\tau_{\rm eff}$ shows evidence of a narrow feature (see, e.g., \S \ref{fits and comparisons}), these overall trends are mostly irrelevant and we will only consider the statistical errors estimated as above.
Nonetheless, it is important to gain some idea of how uncertain the overall redshift evolution we report really is.
We explore this question in this section.\\ \\
We choose to present the calculation with the metal correction based on the \cite{2003ApJ...596..768S} results as the fiducial model. 
The systematic error budget will account for the fact that the correct correction may be closer to that based on the \cite{2005MNRAS.360.1373K} results.

\begin{figure*}[ht]
\begin{center}
\includegraphics[width=1.0\textwidth]{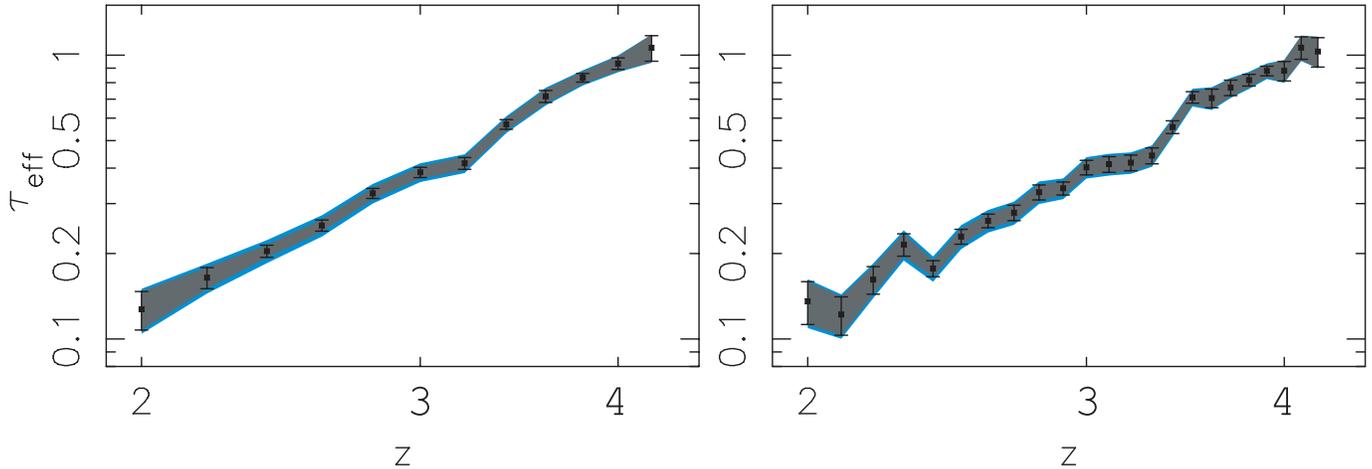}
\end{center}
\caption{Illustration of the systematic error budget associated with the continuum and metal corrections.
The black points show our $\tau_{\rm eff}$ measurements, corrected for continuum bias (\S \ref{continuum estimation}) and with metals removed based on the results of \cite{2003ApJ...596..768S} (\S \ref{metal absorption}), with the corresponding statistical errors calculated as in \S \ref{error bars}.
The dark gray bands include, in addition, the systematic contribution to the error budget from the uncertainty on the continuum correction.
The light blue bands indicate the total error budget, with the uncertainty on the metal correction included as well.
The bands serve to indicate the correlated nature of the systematic errors.
The systematic contribution is at most comparable to the statistical error, and much smaller at both the low and high redshift ends.
The left panel shows our measurement in $\Delta z=0.2$ bins and the right panel shows the measurement with $\Delta z=0.1$ bins.}
\label{systematic errors figure}
\end{figure*}

\subsection{Continuum Bias Systematic Error}
\label{continuum bias systematic error}
In Appendix \S \ref{continuum correction uncertainty}, we estimate the uncertainty on $\tau_{\rm eff}$ arising from the continuum correction to $|\Delta \tau_{\rm eff}|/\tau_{\rm eff}^{\rm true}\lesssim0.0044(1+z)^{1.71}$.
As this is a somewhat conservative upper bound estimate, we identify the $1\sigma$ error to half this value:
\begin{equation}
\sigma_{\tau_{\rm eff}, \rm cont}=0.0022(1+z)^{1.71}.
\end{equation}

\subsection{Metal Correction Systematic Error}
\label{metal correction systematic error}
The systematic error on the metal correction is difficult to assess.
The correction based on the manual line identifications by \cite{2003ApJ...596..768S} is really a lower bound, as some lines are likely missed.
This will necessarily occur most frequently at high redshifts, where line blending is most important.
On the other hand, metal lines coincident with absorption already saturated by hydrogen have little to no impact on the effective optical depth \citep[e.g.,][]{2007MNRAS.382.1657K}, so that it is unclear whether the larger fraction of missed lines at higher redshifts actually makes the estimated correction less accurate. 
The statistical correction based on the results of \cite{2005MNRAS.360.1373K} relies on a linear extrapolation from redward of the forest, which may not be exact.
The two methods, however, are largely independent and a reasonably conservative estimate of the error on the correction is obtained by taking the maximum deviation between the estimates from the two methods over the redshift interval probed.
The deviation between the two corrections is defined as
\begin{equation}
\frac{\Delta \tau_{\rm eff,\alpha}}{\tau_{\rm eff,\alpha}-\tau_{\rm eff}}
\equiv
\frac{|\tau_{\rm eff,\alpha,Schaye}-\tau_{\rm eff,\alpha,Kirkman}|}{\tau_{\rm eff,\alpha,Schaye} - \tau_{\rm eff}},
\end{equation}
where the subscripts `Schaye' and `Kirkman' refer to the measurements corrected for metal absorption as in equations \ref{kirkman correction redshift} and \ref{kirkman correction wavelength}, respectively, and $\tau_{\rm eff}$ is the continuum-corrected measurement prior to metal correction. 
It is largest in the lowest and highest redshifts bins, where it attains $60$\%.\footnote{The one exception is the $z=2.1$ bin for the $\Delta z=0.1$ measurement, for which the deviation is 73\%. We regard this highest value as likely originating from a statistical fluctuation.}
We thus estimate the error contribution of the metal correction to the uncertainty on $\tau_{\rm eff}$ as
\begin{equation}
\sigma_{\tau_{\rm eff}, \rm metal} = 0.6(1 - \tau_{\rm eff,\alpha,Schaye}/\tau_{\rm eff}),
\end{equation}
where the ratio $\tau_{\rm eff,\alpha,Schaye}/\tau_{\rm eff}$ of the metal-corrected measurement to the value prior to correction is given by equation \ref{schaye metal correction}.

\subsection{Total Error}
\label{total error}
The total systematic error budget is obtained by also adding the systematic errors from the continuum and metal correction in quadrature,
\begin{equation}
\sigma_{\tau_{\rm eff}, \rm syst}^{2} = \sigma_{\tau_{\rm eff}, \rm cont}^{2} + \sigma_{\tau_{\rm eff}, \rm metal}^{2}.
\end{equation}
The total uncertainty at each redshift is similarly obtained by adding the statistical contribution estimated as in \S \ref{error bars} and the systematic one in quadrature:
\begin{equation}
\sigma_{\tau_{\rm eff},\rm tot}^{2} = \sigma_{\tau_{\rm eff},\rm stat}^{2} + \sigma_{\tau_{\rm eff},\rm syst}^{2}.
\end{equation}
As discussed above, the systematic term is expected to be strongly correlated in neighboring redshift bins.
The statistical error is estimated as in \S \ref{error bars}, where it was simply denoted $\sigma_{\tau_{\rm eff}}$.\\ \\
The results for both our measurements with $\Delta z=0.1$ and $\Delta z=0.2$ bins are shown in Figure \ref{systematic errors figure}.
The systematic contribution is at most comparable to the statistical error, and much smaller at both the low and high redshift ends.

\section{FITS AND COMPARISONS WITH PREVIOUS WORK}
\label{fits and comparisons}
\subsection{Fits}
\label{fits}
In this section, we fit mathematical functions to our measurements of $\tau_{\rm eff}$ versus redshift, and investigate how well each functional form fits the data.
As we are interested in detecting potential narrow features in the redshift evolution of $\tau_{\rm eff}$, we ignore the slowly varying systematic errors arising from the continuum and metal corrections, and instead consider only the statistical error bars estimated as in \S \ref{error bars}.
We provide our results in graphical and tabular forms.
Theoretical interpretation is left for separate work, with some possibilities outlined in \S \ref{discussion}.\\ \\
\begin{figure*}[ht]
\begin{center}
\includegraphics[width=1.00\textwidth]{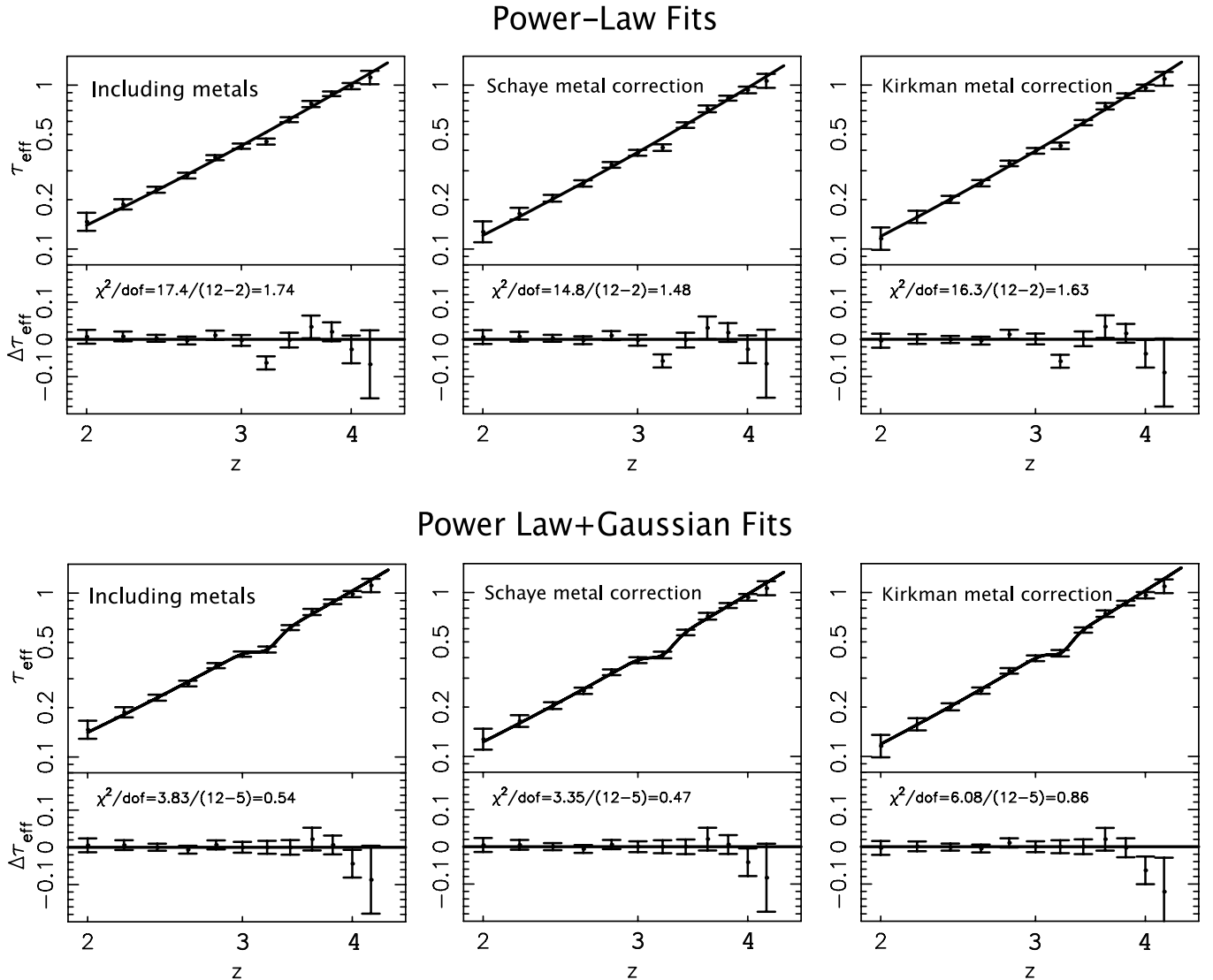}
\end{center}
\caption{Fits to our $\tau_{\rm eff}$ measurements in redshift bins of width $\Delta z=0.2$.
In all cases, the measurements have been corrected for continuum bias as in \S \ref{continuum estimation}.
The top row shows power-law fits and the bottom row shows fits of a power law plus a Gaussian bump.
In the first column, we fit directly to $\tau_{\rm eff}$ (including metal absorption).
In the second and third column, we fit to $\tau_{{\rm eff},\alpha}$, our measurement with metals removed based on the fits of \cite{2003ApJ...596..768S} and on the statistical technique of \cite{2005MNRAS.360.1373K}, respectively.
For each fit, we give the reduced $\chi^{2}$ value, where the number of parameters estimated from the data has been subtracted to obtain the effective number of degrees of freedom in each case.
Best-fit parameters are given in Table \ref{best fits table}.
}
\label{point two fits}
\end{figure*}
\begin{figure*}[ht]
\begin{center}
\includegraphics[width=1.00\textwidth]{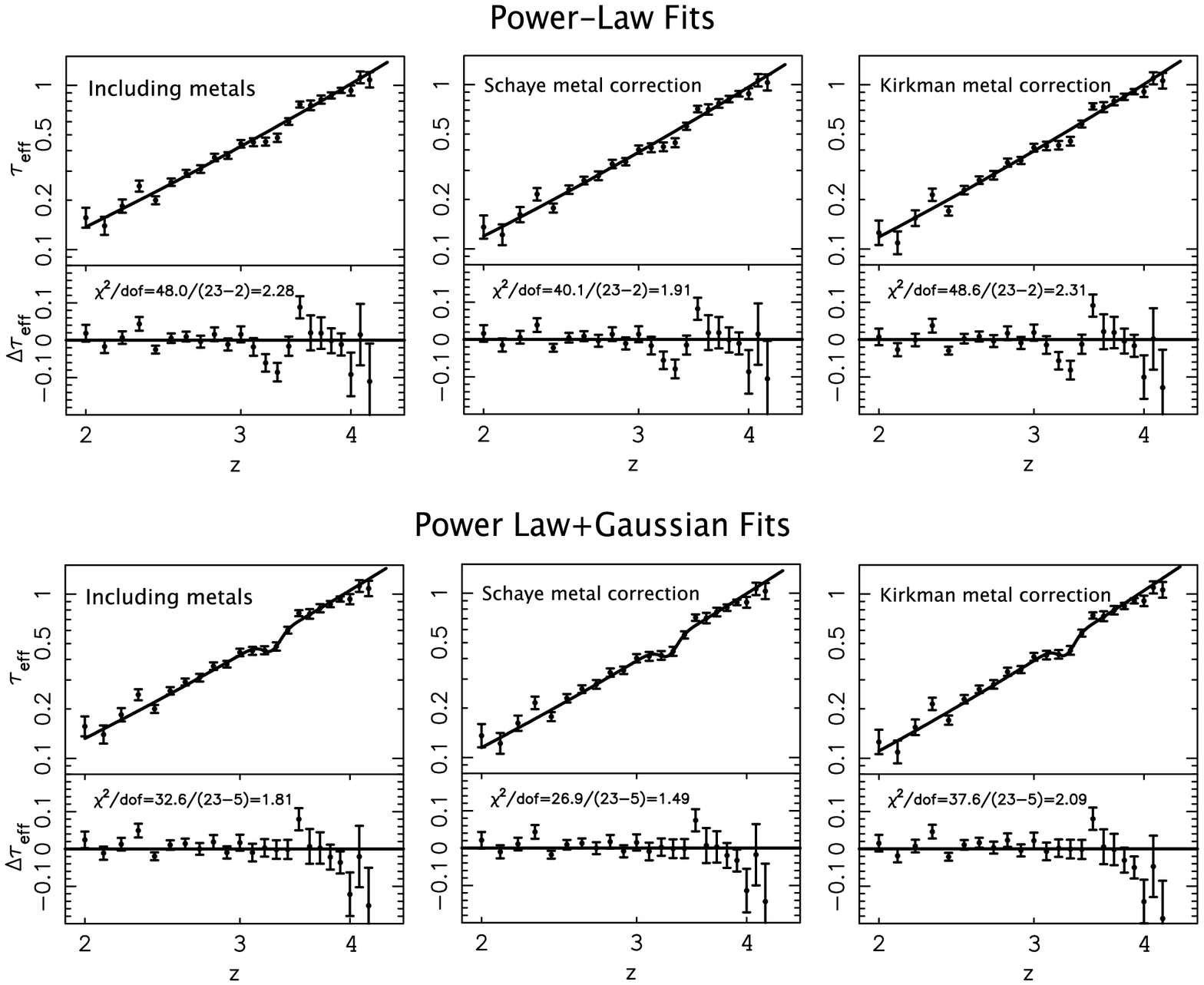}
\end{center}
\caption{Fits to our $\tau_{\rm eff}$ measurements in redshift bins of width $\Delta z=0.1$.
In all cases, the measurements have been corrected for continuum bias as in \S \ref{continuum estimation}.
The top row shows power-law fits and the bottom row shows fits of a power law plus a Gaussian bump.
In the first column, we fit directly to $\tau_{\rm eff}$ (including metal absorption).
In the second and third column, we fit to $\tau_{{\rm eff},\alpha}$, our measurement with metals removed based on the fits of \cite{2003ApJ...596..768S} and on the statistical technique of \cite{2005MNRAS.360.1373K}, respectively.
For each fit, we give the reduced $\chi^{2}$ value, where the number of parameters estimated from the data has been subtracted to obtain the effective number of degrees of freedom in each case.
Best-fit parameters are given in Table \ref{best fits table}.}
\label{point one fits}
\end{figure*}
\begin{deluxetable*}{cccccccccc}
\tablewidth{0pc}
\tablecaption{Parameters for the best-fit power-law and power-law+Gaussian models to our measurements of $\tau_{\rm eff}$, with and without metal corrections, and in redshift bins of width $\Delta z=0.2$ and $\Delta z=0.1$\label{best fits table}}
\tabletypesize{\footnotesize}
\tablehead{
\colhead{} &
\multicolumn{3}{c}{Power law\tablenotemark{a}} &
\multicolumn{6}{c}{Power law+Gaussian\tablenotemark{b}}
\\
\colhead{} &
\colhead{a} &
\colhead{b} &
\colhead{$\chi^{2}/dof$}\tablenotemark{c} &
\colhead{A} &
\colhead{B} &
\colhead{C} &
\colhead{D} &
\colhead{E} &
\colhead{$\chi^{2}/dof$}\tablenotemark{d} \\
\hline \multicolumn{10}{c}{$\Delta z=0.2$}
}
\startdata
Including metals         & -2.706 & 3.880 & 1.74 & 0.00196 & 3.892 & -0.0708 & 4.207 & 0.101 & 0.54 \\
Schaye metal correction  & -2.853 & 4.057 & 1.48 & 0.00141 & 4.062 & -0.0651 & 4.207 & 0.101 & 0.47 \\
Kirkman metal correction & -2.910 & 4.164 & 1.63 & 0.00115 & 4.222 & -0.0673 & 4.218 & 0.0900 & 0.86 \\
\hline \multicolumn{10}{c}{$\Delta z=0.1$} \\ \hline
Including metals         & -2.734 & 3.924 & 2.28 & 0.00153 & 4.060 & -0.0969 & 4.267 & 0.0769 & 1.81 \\
Schaye metal correction  & -2.876 & 4.094 & 1.91 & 0.00111 & 4.225 & -0.0891 & 4.267 & 0.0766 & 1.49 \\
Kirkman metal correction & -2.927 & 4.192 & 2.31 & 0.000876 & 4.403 & -0.0930 & 4.269 & 0.0726 & 2.09
\enddata
\tablenotetext{a}{Model defined in Equation (\ref{power law eq}).}
\tablenotetext{b}{Model defined in Equation (\ref{power law plus gaussian eq}).}
\tablenotetext{c}{We have subtracted the two parameters fitted for from the number of degrees of freedom used to calculated the reduced $\chi^2$.}
\tablenotetext{d}{We have subtracted the five parameters fitted for from the number of degrees of freedom used to calculated the reduced $\chi^2$.}
\end{deluxetable*}

The first functional form that we fit to our measurements is a power law, following previous studies that have found that the evolution of the \Lya~intergalactic opacity was well described by such a function \citep[e.g.,][]{1993ApJ...414...64P}.
Specifically, we solve for the parameters $a$ and $b$ which minimize the $\chi^{2}$ between the model
\begin{equation}
\label{power law eq}
\log \tau_{\rm eff} = a + b\log{(1+z)}
\end{equation}
and the measurements.
These fits to our measurements, corrected for continuum bias and optionally for metal absorption, are shown in the first rows of Figures \ref{point two fits} and \ref{point one fits}, for the data in redshift bins $\Delta z=0.2$ and $\Delta z=0.1$, respectively.
The best-fit parameters and reduced $\chi^{2}$ for each fit are tabulated in Table \ref{best fits table}.
In calculating $\chi^{2}/dof$, we have subtracted the number of parameters fitted for, in this case two, from the number of data points to calculate the number of degrees of freedoms.
For comparison with previous results, the best-fit power law to our continuum-corrected measurement in $\Delta z=0.1$ bins, with metals left in, is
\begin{equation}
\label{tau eff power law fit}
\tau_{\rm eff}=0.0018(1+z)^{3.92} \, .
\end{equation}

For the data with bins of width $\Delta z=0.2$ (Figure \ref{point two fits}), the $z=3.2$ measured $\tau_{\rm eff}$ are seen to lie below the best-fit power laws at the $3\sigma$ level, although the fits globally deviate from pure power laws only at $<2\sigma$.
However, when the data are more finely binned with $\Delta z=0.1$ (Figure \ref{point one fits}), the reduced $\chi^{2}$ for the power-law fits deteriorate substantially, and up to four consecutive points around $z=3.2$ lie below the best-fit power laws.
For example, our measurement corrected for continuum bias but including metals (first panel) has a probability of only 0.0007 of being described by the best-fit power law, with $\chi^{2}/dof=48.0/(23-2)=2.28$.
This is suggestive of a relatively narrow departure from a power law in the redshift evolution of $\tau_{\rm eff}$ near $z=3.2$, which is partially smoothed out when the data are binned in redshift bins of width $\Delta z=0.2$. \\\ \\
In order to investigate the possible departure from a power law around $z=3.2$ in our measurements, and to compare our results with those of \cite{2003AJ....125...32B}, who found similar feature in their precision SDSS measurement, we also fit the same functional form consisting of a power law plus a Gaussian ``bump'' that these authors used,
\begin{equation}
\label{power law plus gaussian eq}
\tau_{\rm eff} = A(1 + z)^{B} + C \exp{\left\{ - \frac{[(1 + z) - D]^{2}}{2 E^{2}} \right\}}.
\end{equation}

These fits are shown in the second rows of Figures \ref{point two fits} and \ref{point one fits} and the best-fit parameters and reduced $\chi^{2}$ for each fit are also tabulated in Table \ref{best fits table}.
When calculating the reduced $\chi^{2}$ for these fits, we have subtracted the 5 parameters fitted for from the $dof$.
For the data with bins of width $\Delta z=0.2$ (Figure \ref{point two fits}), the reduced $\chi^{2}$ are anomalously low ($<1$), suggesting, as may have been expected, that the fitted model allows for too much flexibility given the number of data points.
For the finer $\Delta z=0.1$ bins (Figure \ref{point one fits}), the reduced $\chi^{2}$ are somewhat improved over the pure power-law fits.
For example, it goes down from 2.28 to 1.81 (with corresponding $p$-values of $7\times10^{-4}$ and 0.012, respectively) for the measurement including metal absorption.
However, even after the addition of the Gaussian bump near $z=3.2$ to the power law, the data are not very well described by Equation (\ref{power law plus gaussian eq}).
Indeed, the $z=2.3$ data points exceed the best-fit models at the $3\sigma$ level and, although the evidence is weak owing to the large error bars, there is a hint of a roll over in the $\tau_{\rm eff}$ evolution at $z\gtrsim4$.
For instance, the measurement corrected for metal absorption statistically based on the results of \cite{2005MNRAS.360.1373K} (last panel) yields $\chi^{2}/dof=37.6/(23-5)=2.01$, corresponding to a $p$-value of 0.004. \\ \\
Inspection of Figure \ref{point one fits} suggests that other functional forms may formally fit our data equally well as the power law+Gaussian model with a $z\approx3.2$ downward feature.
This owes to the fact that the level of the error bars on our measurement allow for only a marginal detection of a departure from a pure power law on its basis alone.
In particular, it does not clearly suggest the correct form of the deviation.
The higher precision (but potentially affected by different systematic effects) measurement of \cite{2003AJ....125...32B} did however clearly delineate a downward departure centered at $z\approx3.2$, motivating us to fit the same functional form as these authors.
Moreover, the redshift location and direction of the departure in this model are in qualitative agreement with the expectations of HeII reionization (see \S \ref{introduction}).
Nevertheless, we wish to emphasize that our data alone are unlikely to \emph{require} this particular functional form.
As there is no reason that $\tau_{\rm eff}$ should follow a power law even in the absence of a narrow feature, the parameters of the present fits do not have particular physical significance.
Rather, our fits are merely a convenient device to test for the presence of a departure from a pure power law with an implicit prior that the departure should be similar to the one previously detected by \cite{2003AJ....125...32B}.

\subsection{Comparisons with Previous Work}
\label{comparisons section}
Further insight on the authenticity of the $z\approx3.2$ feature suggested by the data is obtained by comparison with previous measurements.
We compare with our measurement including absorption from metals, since many previous studies did not attempt to remove them.

The existing $\tau_{\rm eff}$ measurements fall in two broad classes: (1) measurements from high-resolution spectra, for which the continuum level of each spectrum is directly estimated on an individual basis, and (2) measurements from low-resolution spectra, with the continuum level estimated from an extrapolation from the red side of the \Lya~line, where there is no absorption from the \Lya~forest.
We begin by briefly describing the measurements of each class that overlap in redshift with ours.\\ \\
\subsubsection{Descriptions of Existing $\tau_{\rm eff}$ Measurements}
For the direct measurements from high-resolution spectra, we focus on those that were obtained from $\gtrsim20$ spectra and for which the evolution with redshift of $\tau_{\rm eff}$ has been estimated.

\cite{2003ApJ...596..768S} manually estimated the continuum level of 14 UVES and 5 HIRES spectra, in a manner similar to ours, with quasar emission redshifts $2.145\leq z_{em}\leq 4.558$.
As discussed in \S \ref{metal absorption}, they repeated their measurement with and without removing metal lines; here we consider their measurement with metal absorption included. 
They did not apply an explicit correction for redshift-dependent continuum bias.

\cite{2004AJ....127.2598S} studied 25 HIRES spectra with emission redshifts $2.31\leq z_{em}\leq 4.10$ and 25 ESI spectra with emission redshifts $4.17\leq z_{em}\leq 6.39$.
She manually estimated the continuum levels for the quasars with $z_{em}\leq4.5$, but extrapolated from the red side of the \Lya~line for the higher-redshift ones.
For the highest-redshift quasars in her sample, she used a fixed power-law spectral index $\alpha=-1.25$ ($f_{\nu}\propto\nu^{-\alpha}$) and set the normalization using the region of the spectra near the rest-frame wavelength 1350~\AA.
She did not apply a redshift-dependent continuum correction.

\cite{2005MNRAS.360.1373K} measured the flux decrement $DA$ at $1.6\leq z_{\Lya} \leq3.2$ using 24 HIRES spectra.
Following the earlier work of \cite{2004ApJ...617....1T} at $z\approx1.9$, they had a team of undergraduate students continuum fit both their actual spectra and mock spectra with matching properties.
They estimated a small bias as a function of $DA$, which they used to correct their measurement.
The maximum correction was $\approx1\%$.
At $z=3$, our continuum bias correction $\approx4\%$ is larger than their maximum correction, although our results are not necessarily inconsistent.
In fact, \cite{2005MNRAS.360.1373K} did not estimate a correction as a function of redshift and most of their data is at $z<3$.
Instead of using a cosmological simulation to generate their mock spectra, \cite{2005MNRAS.360.1373K} randomly placed Voigt profiles, following the line distributions of \cite{1997ApJ...484..672K} and with random redshifts.
These mock spectra thus neglect correlations between the absorption lines and do not represent the true cosmic structure.
\cite{2005MNRAS.360.1373K} find that these mock spectra contain $\approx 5\%$ more total absorption than their real ones.
These differences may thus explain discrepancies that may (or may not) exist between their continuum correction and ours. 
As discussed in \S \ref{metal absorption}, \cite{2005MNRAS.360.1373K} statistically estimated and subtracted the absorption arising from metals in the \Lya~forest using line lists published by \cite{1988ApJS...68..539S}.

\cite{2007ApJ...662...72B} most recently measured the probability density function (PDF) of the transmitted flux $F$ in the \Lya~forest of 55 HIRES spectra, with absorption redshifts spanning $1.7\leq z_{\Lya} \leq 5.8$.
At $z_{\Lya}\leq5.4$, they used a manual spline fit to estimate the quasar continua.
They noted that the transmitted regions at $z_{\Lya}\leq5$ rarely, if ever, reached the continuum, and so their fits were of very low order for these redshifts.
They used a power-law fit to the continuum on the red side of the \Lya~line with $\alpha=0.5$ for quasars with $z_{em}\geq 5.7$.
\cite{2007ApJ...662...72B} removed all the metal lines that could be identified either by damped \Lya~absorption or from multiple metal lines at the same redshift.
They did not, however, mask weak lines found in the forest without counterparts redward of \Lya~emission, with the motivation that doing so would preferentially discard pixels with low \Lya~optical depth (where the metal lines can be seen), introducing a potentially larger bias in the PDF than the one incurred by leaving the contaminated pixels in the sample.
From lognormal fits to the observed optical depth PDF, \cite{2007ApJ...662...72B} inferred the corresponding $\tau_{\rm eff}$ redshift evolution, but did not attempt to estimate the uncertainty in this derived quantity.
They did not apply any continuum correction in this calculation, but noted that if the functional form of the gas density PDF of \cite{2000ApJ...530....1M} is instead assumed, then a continuum correction similar in magnitude to ours must be applied in order to obtain a good fit.

Other works that have directly estimated the transmission of the \Lya~forest from high-resolution spectra, but using smaller data sets, include \cite{1987ApJ...313..171S}, \cite{1995AJ....110.1526H}, \cite{1997ApJ...489....7R}, \cite{2000ApJ...543....1M}, and \cite{2001A&A...373..757K}.
To compare our study to previous analyses of the \Lya~forest from high-resolution spectra, we note that the two previous studies which included the largest samples of quasars over the redshift range we cover are those of \cite{2004AJ....127.2598S} and \cite{2007ApJ...662...72B}.
Thirty of the quasars analyzed by \cite{2004AJ....127.2598S} have a median \Lya~absorption redshift between 2 and 4, where our measurement is most sensitive (Figure \ref{z hists}).
Thirty-nine of the quasars analyzed by \cite{2007ApJ...662...72B} contribute a point with mean \Lya~absorption redshift in this interval.
Our measurement from 86 quasars thus represents an improvement over previous analyses of the \Lya~forest from high-resolution spectra at $2\lesssim z\lesssim 4$ of a factor of about two in the size of the data set alone. 

The measurements using low-resolution spectra, with extrapolation of the continuum from the red side of the \Lya~line, were pioneered by \cite{1993ApJ...414...64P}.
These authors analyzed 29 quasars from the sample of 
\cite{1991AJ....101.2004S} with a resolution of 25~\AA. They extrapolated a continuum of the form $C_{1/2}\lambda^{1/2} + C_{1}\lambda$, where $C_{1/2}$ and $C_{1}$ are parameters that are fit for, accounting for the presence of emission lines redward of \Lya~emission.
They ignored possible features, such as emission lines, in the \Lya~forest region.
They fitted a power law of the form $\tau_{\rm eff}=A(1+z)^{\gamma+1}$ to their data over the redshift range $2.5\leq z_{\Lya} \leq 4.3$, obtaining 
$A=0.0175 - 0.0056\gamma \pm 0.0002$ and $\gamma=2.46\pm0.37$.

\cite{2003AJ....125...32B} analyzed 1061 spectra from the SDSS and from those measured the evolution of $\tau_{\rm eff}$ over the redshift range $2.5\leq z_{\Lya} \leq4$.
They fit a continuum consisting of a power law plus three emission lines (\Lya, 1073~\AA, and 1123~\AA) in the \Lya~forest region, fiducially normalizing each spectrum by its flux in the rest-frame wavelength range 1450-1460~\AA.
They found that the $\tau_{\rm eff}$ evolution was well-fitted by a power law with $\gamma=3.8\pm0.2$, except for a clear feature taking the form of a ``dip'' near $z=3.2$.

\cite{2006ApJS..163...80M} performed a similar analysis on an expanded sample of 3035 SDSS spectra, but using a principal components analysis to model the quasar continuum shape, and normalizing their measurement using the rest-frame wavelength range 1268-1380~\AA.
Their measurement of the mean transmission $\langle F \rangle(z)$ is given in Figure 20 of \cite{2005ApJ...635..761M} and does not show evidence for a pronounced feature near $z=3.2$, although these authors warn that their data points are correlated, that their error bars may not be accurate, and therefore that their measurement should be used for ``qualitative use only''.

The contribution of metals to absorption in the \Lya~forest was not explicitly subtracted in the \cite{1993ApJ...414...64P}, \cite{2003AJ....125...32B}, and \cite{2005ApJ...635..761M} measurements.
However, extrapolating the continuum from redward of \Lya~emission presumably accounts for absorption by metallic transitions with rest-frame wavelength longer than \Lya~to some level.
\\ \\
\begin{figure*}[ht]
\begin{center}
\includegraphics[width=1.00\textwidth]{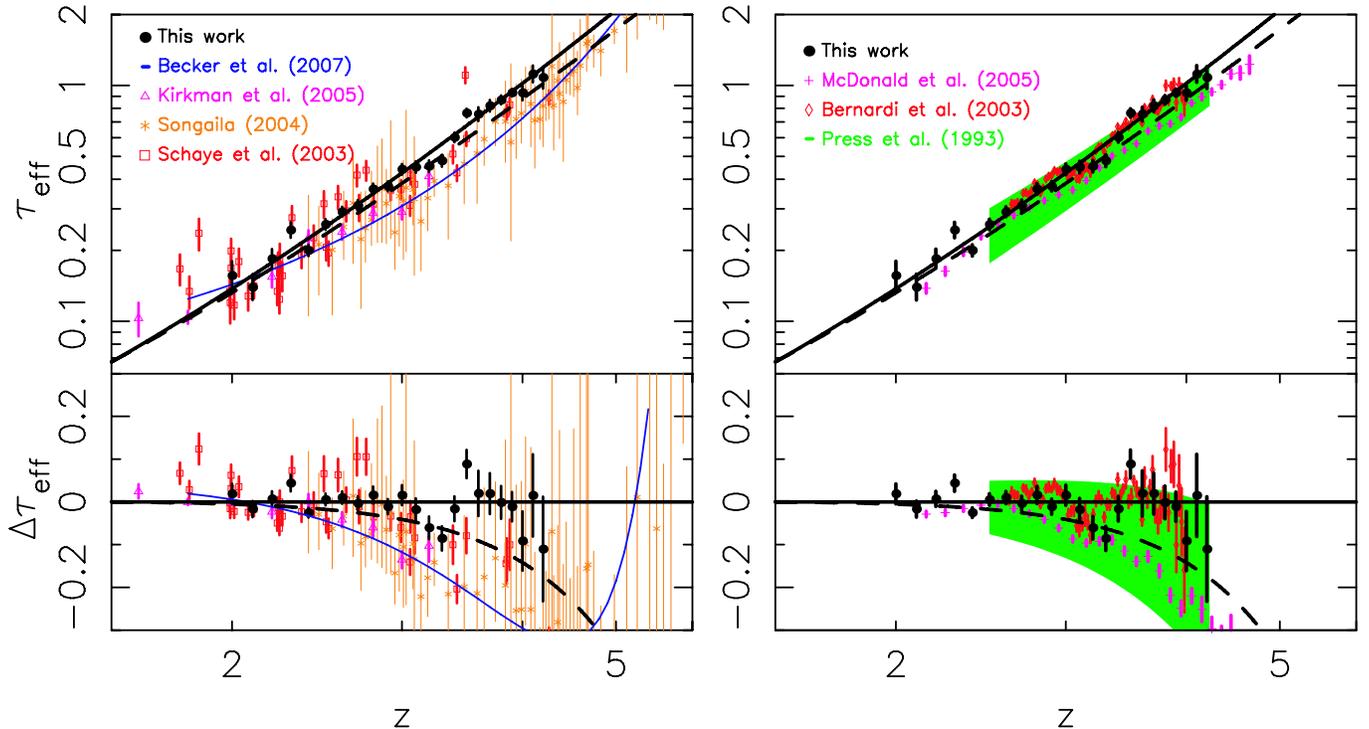}
\end{center}
\caption{Comparisons of our $\tau_{\rm eff}$ measurements with published results.
The left panel compares measurements in which the quasar continuum levels were estimated directly from high-resolution, high-signal-to-noise spectra, as in this work.
The right panel shows estimates based extrapolating the quasar continuum from redward of \Lya~emission in low-resolution spectra.
The solid black curves show the best-fit power law to our measurement corrected for continuum bias (\S \ref{continuum estimation}), but including metal absorption.
Residuals are given with respect to these curves.
The dashed black curves indicate the best-fit power law to our measurement prior to correction for continuum bias.
The published $z\gtrsim3$ points are in better agreement with our measurement prior to correction for continuum bias, as expected since \cite{2003ApJ...596..768S}, \cite{2004AJ....127.2598S}, and \cite{2007ApJ...662...72B} did not apply similar corrections.
Note that none of the published data points for the direct measurements exceeds our best-fit power law near $z=3.2$ (even prior to continuum correction), consistent with the apparent feature at this redshift in our measurement.
Our measurement corrected for continuum bias is seen to trace the SDSS statistical measurement of \cite{2003AJ....125...32B}, who have first claimed detection of a feature at $z=3.2$, remarkably well.
We make this comparison quantitative in Figure \ref{compare bernardi}.
The SDSS measurement of \cite{2005ApJ...635..761M} deviates significantly from ours at $z\gtrsim3$, but these authors warn that their points are correlated and that their error bars may be underestimated.
Our measurement agrees well with the best-fit power law of \cite{1993ApJ...414...64P}, although the uncertainty on the latter is large.
}
\label{comparisons}
\end{figure*}
\subsubsection{Comparisons with Other Direct Measurements}
The published $z\gtrsim3$ points for the direct measurements, shown in the left panel of Figure \ref{comparisons}, are in better agreement with our measurement prior to correction for continuum bias, as expected since \cite{2003ApJ...596..768S}, \cite{2004AJ....127.2598S}, and \cite{2007ApJ...662...72B} did not apply similar corrections.
Note that none of the published data points for the direct measurements exceeds our best-fit power law near $z=3.2$ (even prior to continuum correction), consistent with the apparent feature in at this redshift in our measurement.
We note, in passing, that the error bars on the \cite{2003ApJ...596..768S} points appear underestimated, based on comparing them with the scatter between the individual data points.
Similarly, the error bars on the \cite{2004AJ....127.2598S} points appear overestimated.

The $\tau_{\rm eff}$ curve inferred by \cite{2007ApJ...662...72B} appears to qualitatively differ from our measurement.
These authors did not, however, attempt to directly measure $\langle F \rangle$ or $\tau_{\rm eff}$.
Instead, they assumed that the redshift dependence of the logarithmic mean $\mu$ and standard deviation $\sigma$ of their lognormal fits to the flux PDF was well approximated by a linear function in $1+z$.
From these best-fit linear functions, they calculated $\langle F \rangle(z)$, and hence $\tau_{\rm eff}(z)$.
The shape of their inferred $\tau_{\rm eff}$ curve reflects the assumption of linearity of $\mu$ and $\sigma$ in $1+z$, which may not be exact, especially if the evolution of $\tau_{\rm eff}$ has a feature near $z=3$.
Their measurement is also expected to lie somewhat below ours, since these authors did not apply a continuum correction to the measurement used to produce their $\tau_{\rm eff}$ curve.
\cite{2007ApJ...662...72B} did not quantify the uncertainty on this derived quantity, but it is expected to be significantly larger than the errors on our measurement (G. Becker, private communication), in part because of the smaller number of spectra contributing to the redshift range $2\leq z_{\Lya}\leq 4.2$ in their data set.
A small fraction of the difference between our measurement and the \cite{2007ApJ...662...72B} curve is attributable to the fact that these authors removed most of the metal lines in their measurement, whereas we left them in for the measurement plotted in Figure \ref{comparisons}.
\\ \\
\begin{figure*}[ht]
\begin{center}
\includegraphics[width=0.80\textwidth]{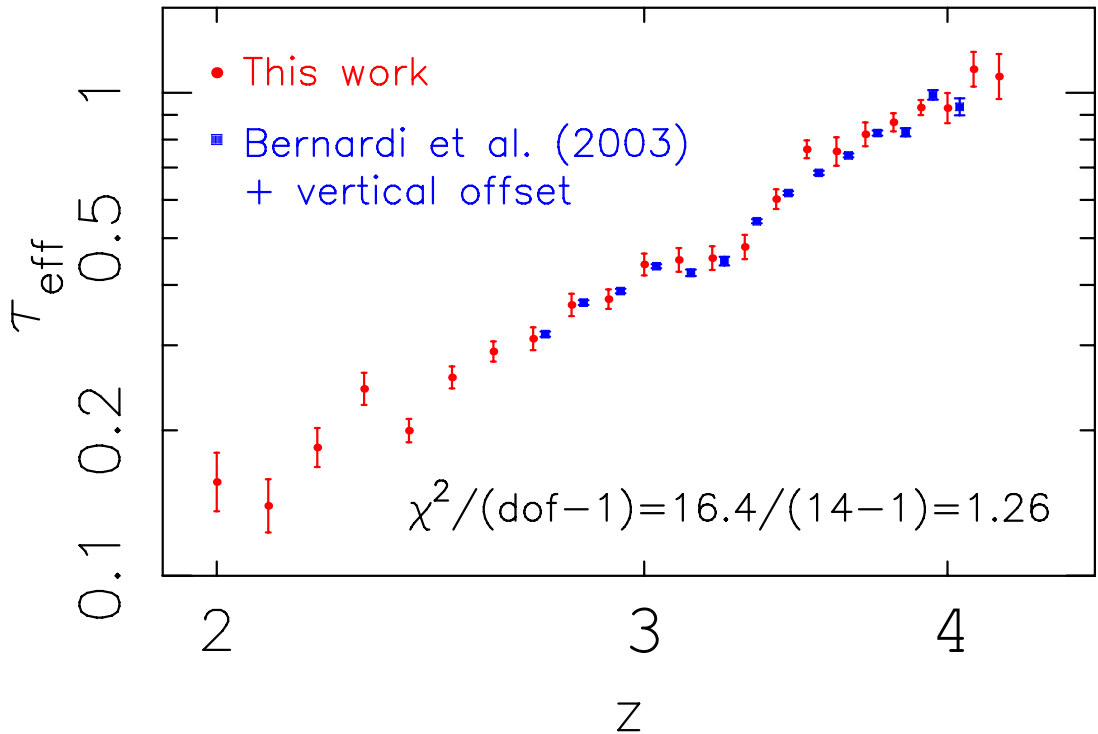}
\end{center}
\caption{Comparison of our measurement of $\tau_{\rm eff}$ corrected for continuum bias (see \S \ref{continuum estimation}; red circles) and of the \cite{2003AJ....125...32B} measurement (blue squares).
The \cite{2003AJ....125...32B} points were binned in the same $\Delta z=0.1$ redshift bins as our measurement.
The blue data points are centered on the same redshifts as the red ones, but have been slightly offset to the right for graphical clarity.
The error bars on the binned \cite{2003AJ....125...32B} points are slightly underestimated, since their neighboring published data points are slightly correlated.
We have added a constant vertical offset in logarithmic space to the \cite{2003AJ....125...32B} points that minimizes the $\chi^{2}$ between the two data sets ($\Delta \log{\tau_{\rm eff}}=0.011$), since their statistical measurement may not be accurately normalized.
Accordingly, we have subtracted one from the number of degrees of freedom used to calculate the reduced $\chi^{2}$.
The resulting $\chi^{2}/dof=16.4/(14-1)=1.26$ indicates that the two data sets agree very well, with a $p$-value of 23\%.}
\label{compare bernardi}
\end{figure*}
\subsubsection{Comparisons with Continuum Extrapolation Measurements}
Our measurement corrected for continuum bias is seen to trace the SDSS statistical measurement of \cite{2003AJ....125...32B}, who first claimed detection of a feature at $z=3.2$, remarkably well.
In Figure \ref{compare bernardi}, we quantitatively compare our measurement of $\tau_{\rm eff}$ corrected for continuum bias with that of \cite{2003AJ....125...32B}.
To do so, we binned the \cite{2003AJ....125...32B} points in the same $\Delta z=0.1$ redshift bins as our measurement.
The error bar on the binned points is estimated as the quadratic sum of the data points that fall in a particular bin, divided by the square root of the number of points in the bin.
These error bars are slight underestimates, since neighboring data points published by \cite{2003AJ....125...32B} are slightly correlated.
The true error bars would only render our conclusion stronger.
We have added a constant vertical offset in logarithmic space to the \cite{2003AJ....125...32B} points that minimizes the $\chi^{2}$ between the two data sets ($\Delta \log{\tau_{\rm eff}}=0.011$), since their statistical measurement may not be accurately normalized\footnote{The normalization on the $\tau_{\rm eff}$ measurement of \cite{2003AJ....125...32B} was obtained by normalizing each spectrum by its flux in the rest-wavelength range 1450-1470~\AA. As these authors show, their absolute $\tau_{\rm eff}$ measurement depends somewhat on the rest-wavelength range used for the normalization.}.
Accordingly, we have subtracted one from the number of degrees of freedom used to calculate the reduced $\chi^{2}$.
The resulting $\chi^{2}/dof=16.4/(14-1)=1.26$ indicates that the two data sets agree very well, with a $p$-value of 23\%.
The $z\approx3.2$ feature discovered by \cite{2003AJ....125...32B} has sometimes been suspected of being an artifact of the SDSS spectrograph, which has a dichroic optical element splitting the beam into blue and red arms near the observed wavelength corresponding to \Lya~at $z\sim3$.
The ESI and HIRES spectrographs on Keck, with which 60 out of the 86 spectra we analyzed were obtained, consist of a single arm and therefore do not suffer from the same potential systematic effect.
Moreover, in this work, we locally estimate the continuum level of each spectrum and so our measurement should be largely unaffected by flux calibration problems, since it is only a function of the normalized transmission. 

The SDSS measurement of \cite{2005ApJ...635..761M} deviates significantly from ours from below at $z\gtrsim3$, although these authors warn that their points are correlated and that their error bars may be underestimated.
It is of note that the \cite{2005ApJ...635..761M} $\tau_{\rm eff}$ measurement lies below even our raw measurement uncorrected for continuum bias, which should be a rather firm lower bound to what $\tau_{\rm eff}$ can be.
Part of the discrepancy may be attributed to the fact that absorption arising from metals present redward of the \Lya~line is implicitly subtracted to some extent in the \cite{2005ApJ...635..761M} estimate.
However, the discrepancy in the highest redshift $z=4.2$ bin is $\sim30\%$, while our estimate of the metal absorption at this redshift is only $2-5\%$ (\S \ref{metal absorption}).
Reconciling the measurements thus appears to require errors on the continuum and metal corrections much larger than the systematic uncertainties estimated in \S \ref{systematic errors}, indicating a genuine discrepancy.

The best-fit power law of \cite{1993ApJ...414...64P} agrees with ours within the uncertainties, although the error on the former is large.

\section{DISCUSSION}
\label{discussion}
We have measured the evolution of the \Lya~effective optical depth over the redshift range $2\leq z\leq4.2$ from a sample of 86 high-resolution, high-signal-to-noise quasar spectra obtained with Keck/ESI, Keck/HIRES, and Magellan/MIKE, paying particular attention to robust error estimation and potential systematic effects.
This represents an improvement over previous analyses of the \Lya~forest from high-resolution spectra in this redshift interval of a factor of two in the size of the data set alone. 

Using mock spectra, we have calculated the redshift-dependent bias on our measurement arising from the increasing cosmological density -- and hence absorption - with redshift (\S \ref{continuum estimation}).
This correction is important at redshifts $z\gtrsim3$, with a magnitude $\approx12\%$ at $z=4$.
Previous direct $\tau_{\rm eff}$ measurements, in which the quasar continua were estimated by fitting the transmission peaks in the \Lya~forest, at these redshifts generally did not apply such a correction and are thus likely to be biased low.
This effect does not affect measurements based on extrapolating the continuum from redward of the \Lya~line, as usually done in measurements based on lower resolution data for which direct local continuum estimation is impossible \citep[e.g.,][]{1993ApJ...414...64P, 2003AJ....125...32B, 2005ApJ...635..761M}.
Such extrapolations, however, likely miss local variations in the continuum from quasar to quasar, making local estimation advantageous for high-resolution data.
As discussed in Appendix \ref{continuum fitting appendix}, the continuum correction estimated in this work is itself subject at some level to systematic uncertainty, arising principally from uncertainties in the cosmological hydrogen photoionization rate, \Gbkg.
This systematic uncertainty could in principle be eliminated by iteratively re-estimating the continuum correction and \Gbkg~from the data itself until convergence is attained.
We defer this involved procedure to future work.

We have also estimated the level of absorption in the \Lya~forest arising from metals, based on results in the literature on both direct and statistical metal removal, finding that this contribution is $\approx6-9\%$ at $z=3$ and decreases monotonically with redshift (\S \ref{metal absorption}).

The high precision of our measurement (attaining $3\%$ in redshift bins with width $\Delta z=0.2$ around $z=3$), indicates significant departures from the best-fit power law evolution ($\tau_{\rm eff}=0.0018(1+z)^{3.92}$ for our measurement corrected for continuum bias, but including metals; Table \ref{best fits table}), particularly near $z=3.2$, where there is evidence of a dip (\S \ref{fits and comparisons}).
Our measurement, with the feature near $z=3.2$, is in excellent agreement with the precision measurement from SDSS spectra of \cite{2003AJ....125...32B} (\S \ref{comparisons section}), who used an independent approach.

\cite{2002ApJ...574L.111T} interpreted the feature seen by \cite{2003AJ....125...32B} as a signature of HeII reionization.
Although the feature is indeed in qualitative agreement with what might be expected from a temperature jump in the IGM owing to HeII reionization, degeneracies exist.
Neglecting redshift-space distortions, the \Lya~optical depth at any point in the forest is given by
\begin{equation}
\label{gp tau}
\tau=
\frac{\pi e^{2} f_{\Lya}}{m_{e} \nu_{\Lya}}
\frac{1}{H(z)}
\frac{R(T)n_{\rm HII}n_{e}}{\Gamma^{bkg}},
\end{equation}
where we have made the approximation of photoionization equilibrium \citep{1965ApJ...142.1633G}.
Here, $e$ is the electron charge, $m_{e}$ is the electron mass, $f_{\Lya}$ is the \Lya~oscillator strength, $\nu_{\Lya}$ is the \Lya~frequency, $R(T)\propto T^{-0.7}$ is the hydrogen recombination coefficient \citep[e.g.,][]{2006agna.book.....O}, $n_{\rm HII}$ is the ionized hydrogen number density, $n_{e}$ is the free electron number density, and $\Gamma^{bkg}$ is the hydrogen photoionization rate.
From Equation (\ref{gp tau}), we see that at least three things could produce a feature in the redshift evolution of the optical depth:
\begin{enumerate}
\item A change in temperature: an upward jump in $T$ will result in a downward jump in $\tau$;
\item a change in the free electron number density: an upward jump in $n_{e}$ will result in an upward jump in $\tau$;
\item a change in the ionizing background: an upward change in $\Gamma^{bkg}$ will result in a downward change in $\tau$.   
\end{enumerate}

Both an upward temperature change and an increased free electron number density could arise from the reionization of HeII, although the latter effect is limited to $7\%$ if HeI has already been reionized (e.g., by stars simultaneous to HI reionization), for a cosmic He mass fraction $Y=0.25$.
A temperature jump of a factor of two \citep{2000ApJ...534...41R, 2003ApJ...596..768S} could suppress $\tau$ by up to $2^{-0.7}\approx40\%$, although redshift-space distortions could mitigate this effect at some level \citep{2002ApJ...574L.111T}.
However, even if HeII reionization is not happening and the IGM temperature is constant, $\tau_{\rm eff}$ could well display a feature if the ionizing background is evolving suitably, for example if it is transitioning from being dominated by stars to being dominated by quasars. 
For instance, empirical studies \citep[e.g.,][]{2007ApJ...654..731H} indicate that the contribution to the ionizing background from quasars peaked at $z\approx 2$, with small uncertainty, but simple theoretical arguments \citep{2003MNRAS.341.1253H} imply that the contribution from stars likely peaked much earlier.
We must note that the effective optical depth we have measured is not directly related to the local optical depth as given by equation \ref{gp tau}, but in fact depends on the PDF of the transmitted flux through its first moment (eq. \ref{tau eff}). 

In addition to the degeneracies outlined above, the theoretical interpretation of the feature we find evidence for near $z=3.2$ poses significant challenges.
Why does the feature appear so narrow?
If it is due to a temperature jump of a factor of two from HeII reionization and the IGM cools mainly adiabatically, then the Universe must expand by a factor of $\sqrt{2}$ in order for the IGM temperature to go back to its pre-jump value.
Yet, the feature appears to have a much smaller width ($\Delta z\sim0.5$ at $z\sim3.2$; e.g., Fig. \ref{point one fits}).
Cosmic variance may also be expected to smooth the observed feature, which is an average over many sightlines, although it is not clear whether this would act to increase or decrease its width.
Moreover, our estimate of the duration of the event may be misled by our perception of $\tau_{\rm eff}$ evolution as a power law plus a narrow downward feature.
Indeed, even in the absence of an event such as HeII reionization near $z=3$, there is apparently no good reason to expect $\tau_{\rm eff}$ to evolve as a pure power law, so that the baseline against which we are comparing the feature may not be correct.
A feature in the redshift evolution of the mean transmission of the \Lya~forest may also be expected to impact its flux power spectrum.
Precision measurements of the large-scale flux power from SDSS \citep[e.g.,][]{2006ApJS..163...80M} apparently do not show a corresponding feature.
Preliminary calculations, however, suggest that the large-scale power need not be affected by a feature in the mean transmission of the forest, if it arises from a change in the temperature of the gas, since a change in the temperature modifies the range of gas densities to which the power spectrum is sensitive in a way that tends to counteract the effect of the change in mean transmission.

We regard the questions of whether HeII reionization is occurring near $z=3.2$, whether the detection of a dip in the evolution of the \Lya~effective optical depth near this redshift could be a signature of this process, or what are the other implications of our precision measurement of the evolution of \Lya~effective optical depth, as rich and complex.
We will investigate these questions theoretically in separate work.

Observationally, additional clues on the evolution of the physical state of the IGM at $2\lesssim z \lesssim 4$ could be provided in the immediate or near future by measurements of higher-order ($\beta$, $\gamma$, ...) Lyman transitions, which probe different density regimes.
Precise measurements of the small-scale flux power spectrum of the forest, which is perhaps most sensitive to the temperature of the IGM, with fine redshift sampling may offer our best opportunity to break the degeneracy between a temperature change and a pure evolution of the hydrogen photoionizing background.

\acknowledgements
We are particularly grateful to Scott Burles and John O'Meara for providing us with the MIKE data used in this work and for useful comments.
We further thank Wallace Sargent and Michael Rauch for allowing us to use some of their HIRES spectra and for useful suggestions.
George Becker kindly provided us with the numerical values used to include his $\tau_{\rm eff}$ measurement in Figure \ref{comparisons}.
We thank Suvendra Dutta for assistance with the simulation used to construct the mock spectra and acknowledge useful discussions with Matthew McQuinn and Patrick McDonald.
Some of the data presented herein were obtained at the W.M. Keck
Observatory, which is operated as a scientific partnership among the
California Institute of Technology, the University of California and the
National Aeronautics and Space Administration. The Observatory was made
possible by the generous financial support of the W.M. Keck Foundation.
The rest of the data analyzed in this work were gathered with the 6.5 meter Magellan Telescopes located at Las Campanas Observatory, Chile.
CAFG is supported by a NSERC Postgraduate Fellowship.
JXP is supported by an NSF CAREER grant (AST-0548180) and by 
NSF grant AST-0709235.
This work was further supported in part by NSF grants ACI
96-19019, AST 00-71019, AST 02-06299, AST 03-07690, and AST 05-06556, and NASA ATP grants NAG5-12140, NAG5-13292, NAG5-13381, and NNG-05GJ40G.
Additional support was provided by the David and Lucile Packard, the Alfred P. Sloan, and the John D. and Catherine T. MacArthur Foundations.

\appendix
\section{Tests for Systematic Effects}
We present in this section tests that we performed to ensure that our measurement is free of significant systematic errors.

\subsection{Different Instruments}
\begin{figure*}[ht]
\begin{center}
\includegraphics[width=0.80\textwidth]{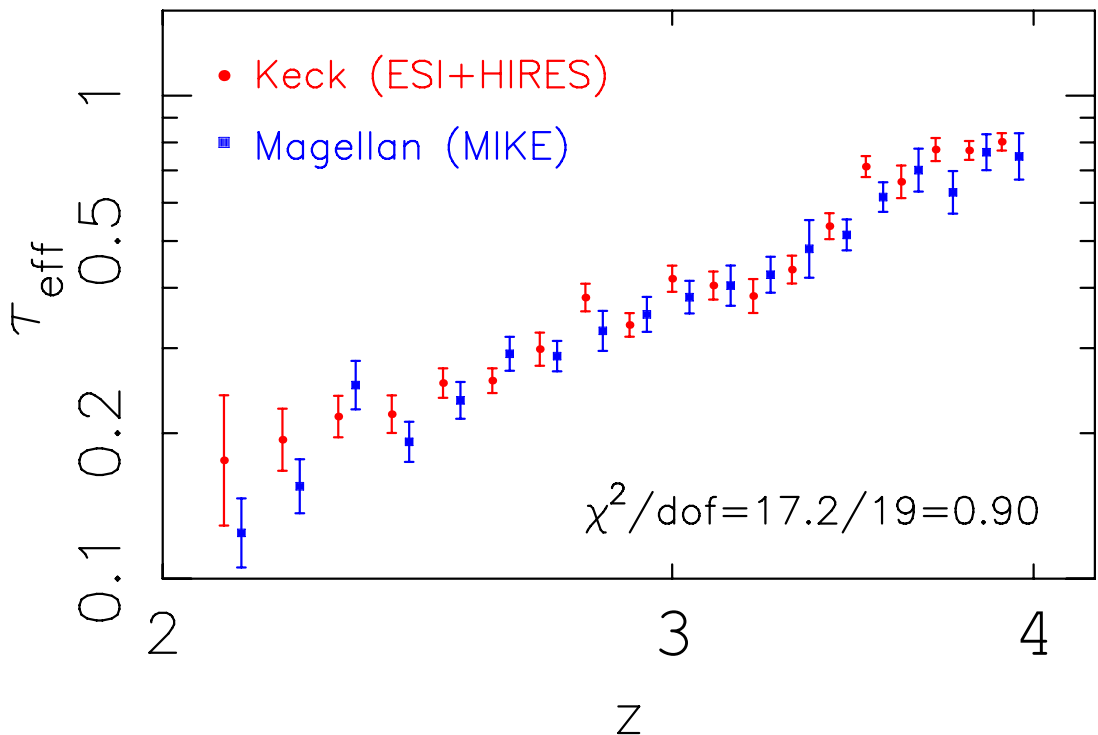}
\end{center}
\caption{Comparison of $\tau_{\rm eff}$ (uncorrected for continuum bias and with metals left in) as measured from the Keck (HIRES+ESI; red circles) and Magellan (MIKE; blue squares) spectra only.
The blue data points are centered on the same redshifts as the red ones, but have been slightly offset to the right for graphical clarity.
The reduced $\chi^{2}=0.90$ between the two measurements indicate that the data from the different observatories give perfectly consistent results.
}
\label{compare datasets}
\end{figure*}
As detailed in \S \ref{data set}, the spectra we analyzed in this work were obtained with the HIRES and ESI spectrographs on Keck, and with the MIKE spectrograph on Magellan.
To verify that the data obtained with the different instruments give consistent results, we have repeated our raw $\tau_{\rm eff}$ measurement (uncorrected for continuum bias and with metals left in), separately using only data from Keck (HIRES+ESI), then only data from Magellan (MIKE).
We have chosen to group the HIRES and ESI data into Keck data because the small number of HIRES spectra (16, distributed as illustrated in Figure \ref{z hists}) makes it difficult to calculate reliable error bars on the measurement from these spectra alone.
The result of this comparison is shown in Figure \ref{compare datasets}.
The reduced $\chi^{2}=0.90$ between the two measurements indicate that the data from the different observatories give perfectly consistent results.
Figure \ref{spectro scatter} in \S \ref{continuum estimation} provides a similar test, combining the data from different redshift bins by normalizing them according to the mean transmission in the bins. 

\subsection{Ly$\beta$-OVI Emission and Associated Absorption}
\begin{figure}[ht]
\begin{center}
\includegraphics[width=1.0\textwidth]{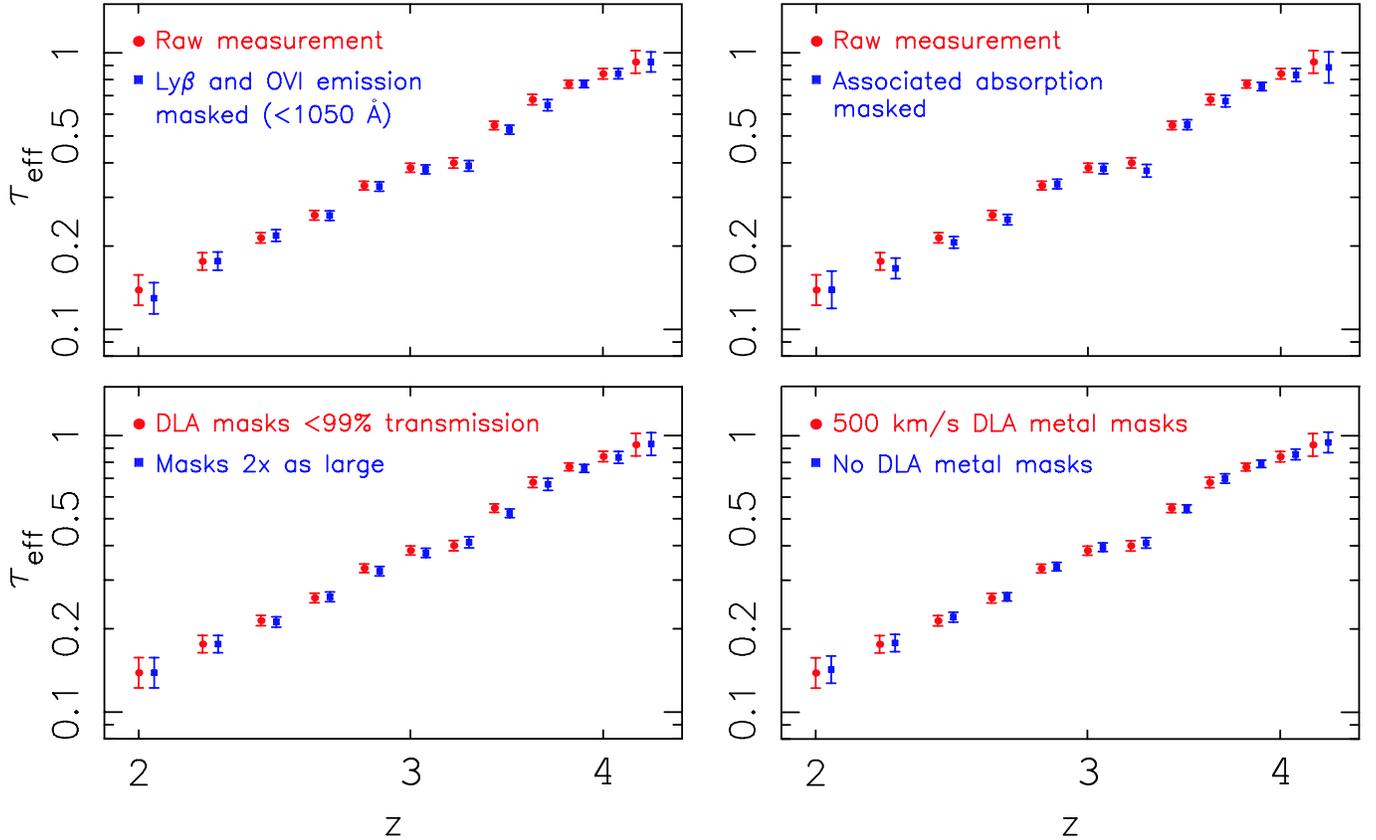}
\end{center}
\caption{Tests for systematic effects associated with choices of analysis parameters.
In each panel, the red circles show our raw $\tau_{eff}$ measurement (with metals left in and uncorrected for continuum bias) and the blue squares show the corresponding measurement for a variation of the analysis parameters.
\emph{Top left} (Ly$\beta$ and OVI emission): Rest-frame wavelengths $<1050$~\AA~masked.
\emph{Top right} (associated absorption): Metal lines with redshift within 3000 km s$^{-1}$ of the quasars masked.
\emph{Bottom left} (width of SLLS and DLA masks): Masks around SLLS and DLAs twice as large as the fiducial width (SLLS or DLA transmission $<99\%$).  
\emph{Bottom right} (masks on metal lines associated with SLLS and DLAs): Not masking metal lines associated with SLLS and DLAs.
The blue points are centered on the same redshifts as the red ones, but have been slightly offset to the right for graphical clarity.
In each panel, the two measurements agree within 1$\sigma$ over the entire redshift range covered, indicating that our results are robust to the choices of analysis parameters.
}
\label{systematics}
\end{figure}

Some authors \citep[e.g.,][]{2007ApJ...662...72B} have previously masked a region redward of Ly$\beta$ in their \Lya~forest analyses to avoid potential problems arising from the presence of the Ly$\beta$ emission line itself, as well as from OVI emission at 1038~\AA.
In this work, it was unclear that these emission lines were problematic.
We thus chose to keep all \Lya~forest wavelengths redward of Ly$\beta$, except for the proximity regions of the quasars, in our analysis (\S \ref{forest definition}).
We repeated our raw measurement (with metals left and uncorrected for continuum bias) masking rest-frame wavelengths $<1050$~\AA~and found no significant difference, as shown in the top left panel of Figure \ref{systematics}.\\ \\

Quasar spectra occasionally show evidence for metal absorption associated with the host galaxies.
We tested whether this could affect our measurement by masking metal lines (as for those associated with SLLS and DLAs, as described in \S \ref{masks}) with redshift within 3000 km s$^{-1}$ of our quasars.
The width of these masks was chosen to account for the large typical uncertainties in quasar redshift determinations \citep[for related discussions on quasar redshift uncertainties, see e.g.,][]{1982ApJ...263...79G, 1992ApJS...79....1T, 2001AJ....122..549V, 2002AJ....124....1R, 2006ApJ...651...61H, 2007AJ....133.2222S, 2007astro.ph..1042F}.
As the top right panel of Figure \ref{systematics} shows, masking this potential associated absorption also has no significant effect on our measurement.

\subsection{Super Lyman-Limit Systems and Damped \Lya~Absorbers}
\label{dla tests}
Many of the lines of sight analyzed in this work were originally selected to contain super Lyman-limit systems and damped \Lya~absorbers.
In order to make an unbiased measurement of $\tau_{\rm eff}$, we have masked the systems with $\log{N_{\rm HI}}\geq19.0$, which individually contribute relatively large equivalent widths to the absorption, as described in \S \ref{masks}.
In this section, we justify our claim that the resulting measurement should be nearly unbiased and test for systematic effects that may be introduced by the masks.

We first show that in a random sample of sightlines, the contribution of systems with $\log{N_{\rm HI}}\geq19.0$ to $\tau_{\rm eff}$ is small.
Thus, if these systems are masked from a biased sample, the unmasked regions should have a $\tau_{\rm eff}$ close to that of a random sample.\footnote{There is an implicit assumption in this argument that the masked regions are uncorrelated with the unmasked ones. This assumption is satisfied, since the SLLS and DLA masks typically have widths of tens of proper Mpc, compared to a correlation length of about 1 proper Mpc in the \Lya~forest (see \S \ref{error bars} for a discussion of this correlation length).}

We can estimate the contribution of SLLS and DLAs to the opacity of the \Lya~forest from the observed column density distribution.
Specifically, the average effective optical depth owing to systems with column density in the range [$N_{min}$, $N_{max}$] can be written as
\begin{equation}
\langle \tau_{\rm eff} \rangle= 
\frac{1+z}{\lambda_{\Lya}}
\int_{N_{min}}^{N_{max}}
dN_{\rm HI}
\int db f(N_{\rm HI},~b)W(N_{\rm HI},~b),
\end{equation}
where $f(N_{\rm HI},~b)$ is the joint PDF of the column density and the Doppler $b$-parameter, and $W(N_{\rm HI},~b)$ is the equivalent width of an absorber with column density $N_{\rm HI}$ and a given $b$-parameter \cite[e.g.,][]{1993A&A...278..343Z}.
We set $\log{N_{min}}=19.0$ and $\log{N_{max}}=22.0$.
This maximum was chosen so that systems of column density this high are too rare to make a difference even in our biased sample, owing to the cutoff that is observed in the distribution of DLA column densities \citep[e.g.,][]{2001ApJ...562L..95S}.
We assume for simplicity that $N_{\rm HI}$ and $b$ are statistically independent, and that the $b$-parameter distribution is a $\delta$-function centered on $b=30$ km s$^{-1}$.
As expected, since the equivalent width depends only weakly on the $b$-parameter for the large column densities of interest, setting $b=20$ km s$^{-1}$ or $b=40$ km s$^{-1}$ instead makes no practical difference.
For the column density distribution, we use the \cite{1996ApJ...461...20H} parameterization, in which $f(N_{\rm HI})\propto N_{\rm HI}^{-1.5}$.

\begin{figure}[ht]
\begin{center}
\includegraphics[width=0.80\textwidth]{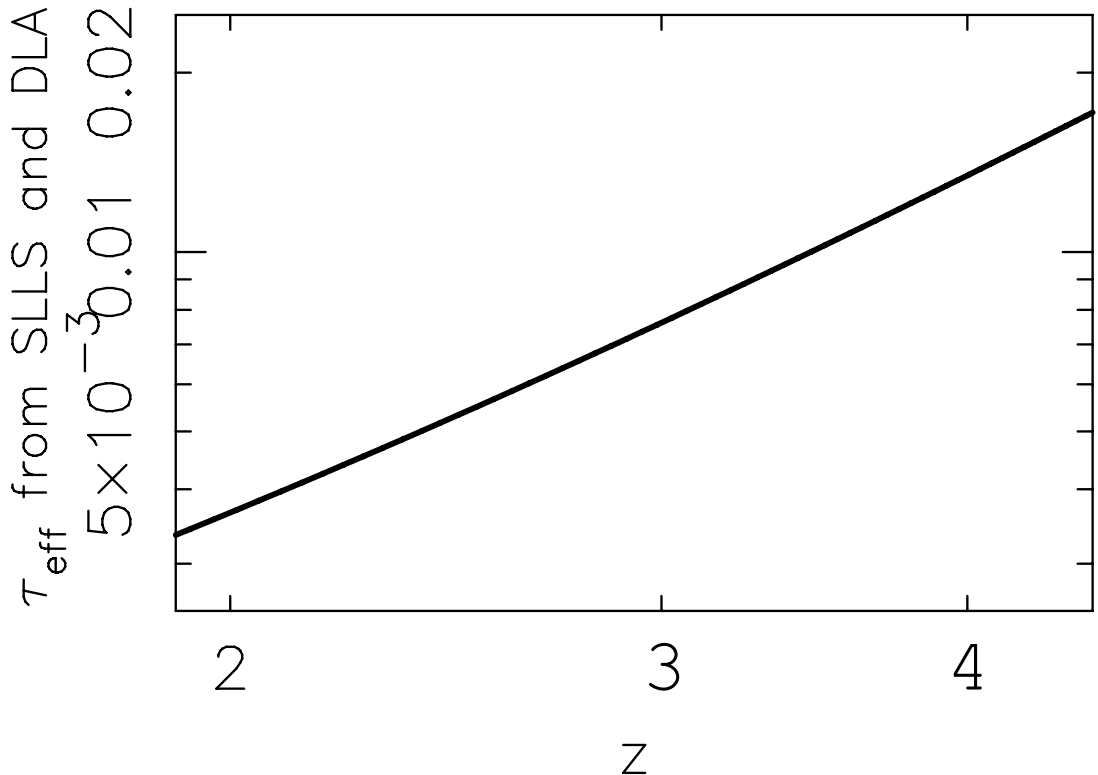}
\end{center}
\caption{Contribution of discrete absorbers with column density $\log{N_{\rm HI}}\geq19.0$ to the total effective optical depth of the \Lya~forest as a function of redshift in an unbiased sample of sightlines.
This contribution is fractionally small ($\lesssim2\%$) over the redshift range covered by our $\tau_{\rm eff}$ measurement.
}
\label{DLA tau eff}
\end{figure}

Figure \ref{DLA tau eff} shows the contribution of discrete absorbers with column density $\log{N_{\rm HI}}\geq19.0$ to the total effective optical depth of the \Lya~forest as a function of redshift in an unbiased sample of sightlines.
This contribution is fractionally small ($\lesssim2\%$) over the redshift range covered by our $\tau_{\rm eff}$ measurement.
Therefore, if these systems are completely masked in a biased sample, $\tau_{\rm eff}$ is underestimated by this small amount with respect to an unbiased sample. 

Although the above argument is in principle sufficient, we have also verified that our results are insensitive to the width of our SLLS and DLA masks, thus confirming that no artificial trends are introduced by them.
Specifically, we compare in the bottom left panel of Figure \ref{systematics} our uncorrected measurement with the same measurement but with masks twice as large, and find the two measurements agree within 1$\sigma$ over the entire redshift range covered.
The bottom right panel of Figure \ref{systematics} shows that the measurement is also insensitive to masking the metal lines associated with the SLLS and the DLAs. 

\section{Continuum Error Details}
\label{continuum fitting appendix}
In this section, we estimate the errors made when continuum fitting quasar spectra manually, as done to measure the evolution of the \Lya~effective optical depth in this paper. \\ \\
\subsection{Variance Between Continuum Fitters}
\begin{figure*}[ht]
\begin{center}
\includegraphics[width=0.80\textwidth]{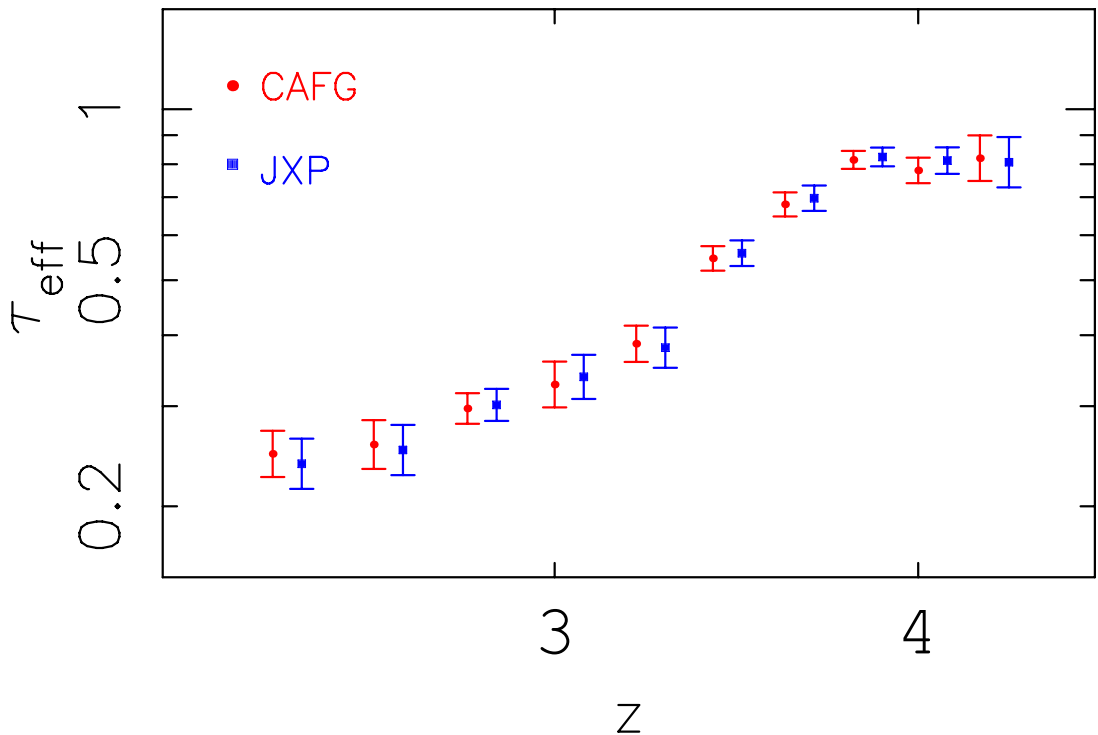}
\end{center}
\caption{Comparison of $\tau_{\rm eff}$ as inferred from a subset of 31 ESI spectra, where the spectra have been continuum fitted independently by CAFG (red circles) and by JXP (blue squares).
The blue data points are centered on the same redshifts as the red ones, but have been slightly offset to the right for graphical clarity.
The $\tau_{\rm eff}$ values from the different continuum fits agree within 2\% on average, and everywhere within $4\%$.
The discrepancies are generally much smaller than the 1$\sigma$ errors on the measurements, indicating that variance in the continuum fits is a negligible source of error. 
}
\label{compare cafg vs jxp}
\end{figure*}

First, we ask how much do continuum fits of the same spectra vary when they are made by different individuals?
A sub-sample of 31 of our ESI spectra were independently continuum fitted by two of us, J.~X. Prochaska and C.-A. Faucher-Gigu\`ere.
The continuum fits were found to agree surprisingly well between the two continuum fitters, as illustrated in Figure \ref{compare cafg vs jxp}.
The $\tau_{\rm eff}$ values from the different continuum fits agree within 2\% on average, and everywhere within $4\%$, for redshift bins of width $\Delta z=0.2$.
The discrepancies are generally much smaller than the 1$\sigma$ errors on the measurements, indicating that variance in the continuum fits is a negligible source of error. 
\\ \\
\subsection{Systematic Bias}
Having determined that the scatter between different continuum fits of the same spectra is small, we turn to the question of whether these may be systematically biased with respect to the true quasar continuum.
Of course, the true continuum level is unknown in actual quasar spectra.
We therefore make use of mock spectra, for which the continuum level is known by construction.

We generate transmission fields $F(\lambda)$ from a cosmological dark matter simulation, as described in detail in \cite{2007astro.ph..1042F}.
Briefly, the baryons are assumed to trace the dark matter, except for Jeans smoothing, and obey a simple power-law temperature-density relation \citep{1997MNRAS.292...27H}.
The redshift-space distortions owing to thermal broadening and peculiar velocities are explicitly modeled.
The simulation box we use has $1024^{3}$ particles with side length $130~h^{-1}$ comoving Mpc.
For the hydrogen photoionizing background, $\Gamma^{bkg}$, we linearly interpolate between the values given by \cite{2005MNRAS.357.1178B} at $2\leq z\leq4$; at $z=4.5$, we interpolate between the \cite{2005MNRAS.357.1178B} $z=4$ value and the $z=5$ value given in \cite{2004MNRAS.350.1107M}.
This results in $\Gamma^{bkg}=(1.3, 1.1, 0.9, 0.95, 1.0, 0.655)\times 10^{-12}$ s$^{-1}$ at $z=(2, 2.5, 3, 3.5, 4, 4.5)$. 

As real quasar spectra are not featureless, but display a number of weak emission lines in \Lya~forest, we multiply our normalized transmission fields by a composite quasar spectrum made from a sample of 657 quasars from the FIRST Bright Quasar Survey \citep[][]{2001ApJ...546..775B}.
The mock spectra are degraded to realistic resolution, pixel size, and signal-to-noise.

Careful manual continuum fitting is a very time-consuming procedure, and so it is impossible to explore the full redshift, resolution, and signal-to-noise parameter space covered by our actual spectra.
The tests of \S \ref{continuum estimation} support that there is little, if any, trend of continuum error with noise and resolution for our high-quality data.
We thus concentrate on the particular case of ESI spectra, for which we have the most data at redshifts $z\gtrsim3$, where we expect the continuum error to be largest, owing to the increasing cosmological matter density.
The ESI spectra are also those with the coarsest resolution used in this work.
Specifically, we set \mbox{$FWHM=40$ km s$^{-1}$}, 11 \mbox{km s$^{-1}$} pixels, and $S/N=20$ per pixel.
We generated 10 different mock spectra with these properties at redshift intervals $\Delta z=0.5$ from $z=2$ to $z=4.5$, and continuum fitted each of them with the same procedure with which we fitted the continua of the real spectra.
As a check, we also continuum fitted a few mock HIRES spectra (\mbox{$FWHM=6$ km s$^{-1}$}, 2 \mbox{km s$^{-1}$} pixels, and $S/N=30$ per pixel), corresponding to the highest resolution spectra in our sample, and confirmed that the continuum correction (see below) for these was approximately the same as for the ESI spectra at $z=4$.
The true mock continuum, while known by construction, was hidden during the continuum fitting.
In generating the mock spectra, we wrapped around our simulation box 6 times for each spectrum, for a total path length of $6\times130~h^{-1}$ comoving Mpc.
Each mock spectrum thus contains more \Lya~forest data than an actual spectrum would between 1025~\AA~and 1216~\AA, by a factor $\approx2$ at $z=3$.
We intentionally did not model the cosmological evolution of the \Lya~forest in a given quasar spectrum versus observed wavelength.
For example, the entire \Lya~forest of one of our $z=4$ mock spectra arises from a $z=4$ density field subject to a $z=4$ photoionizing background.
Therefore, the mean continuum errors that we calculate at a given redshift are valid for this exact redshift.

\begin{figure*}[ht]
\begin{center}
\includegraphics[width=0.95\textwidth]{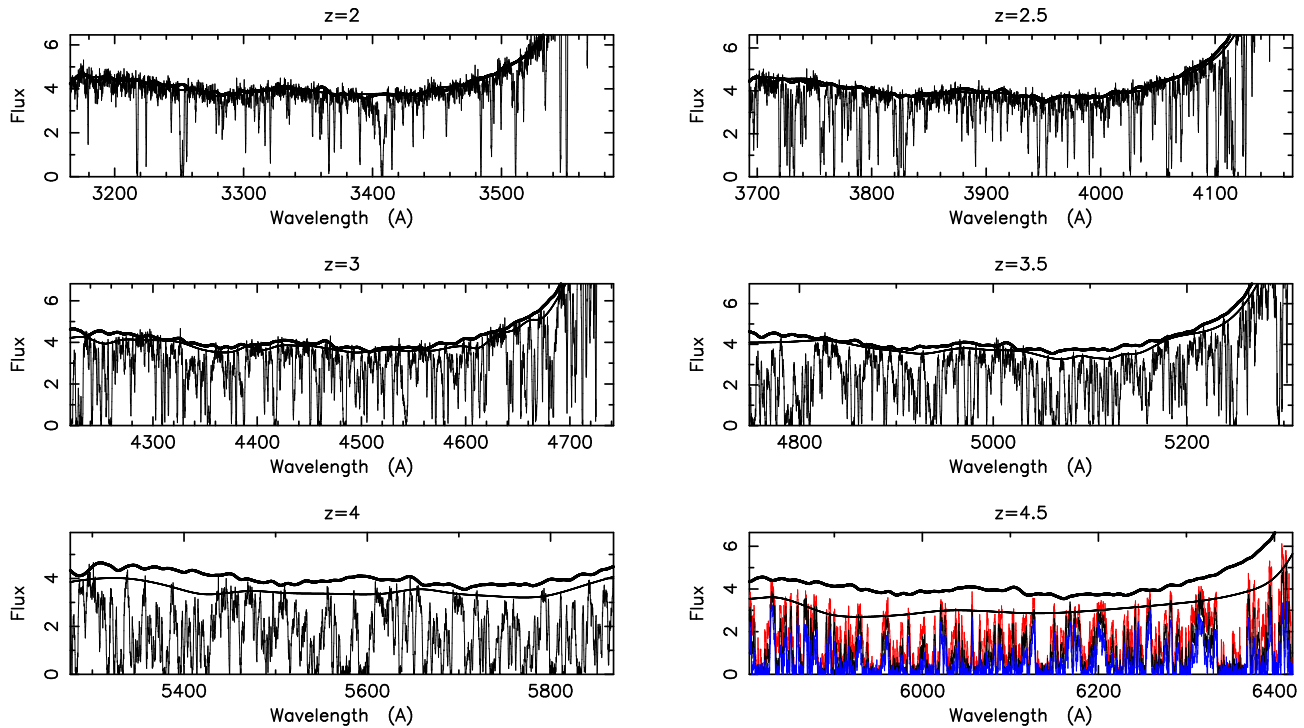}
\end{center}
\caption{Examples of continuum fits to mock quasar spectra.
Mock ESI spectra of quasars from redshift $z=2$ to $z=4.5$ are shown at intervals $\Delta z=0.5$ successively in the panels.
In each case, the thick curve shows the known true continuum level and the blindly estimated continuum is indicated by the thinner horizontal curve.
At $z=2$, the estimated continuum is practically indistinguishable from the true continuum.
The discrepancy between the true and estimated continua increases with redshift, as quantified in Figure \ref{continuum trend}.
At $z=4$, the continuum fits underestimate the true continuum by almost $12\%$ on average.
In the $z=4.5$ panel, we show corresponding spectra, but for the background hydrogen photoionization rate, \Gbkg, half the fiducial value (blue) and twice the fiducial value (red).
}
\label{example cont fits}
\end{figure*}

Figure \ref{example cont fits} shows examples of continuum fits on mock spectra.
At $z=2$, the estimated continuum is practically indistinguishable from the true continuum.
The discrepancy between the true and estimated continua increases with redshift.
At $z=4$, the continuum fits underestimate the true continuum by almost $12\%$ on average.\\ \\

Let us denote by $C_{\rm est}$ the estimated continuum level, by $C_{\rm true}$ the true continuum, and define $\Delta C\equiv C_{\rm true} - C_{\rm est}$.
In Figure \ref{continuum trend}, we plot $\langle \Delta C/C_{\rm true} \rangle$ vs. $z$, where the average is calculated from our 10 continuum fits to mock spectra at each redshift.
The 1$\sigma$ uncertainty in this mean is estimated similarly to the error in our $\langle F \rangle$ measurement (c.f. \S \ref{error bars}).
Specifically, each spectrum was broken in segments of 2000 pixels, resulting in 33 chunks at $z=2$ and 42 at $z=4.5$.
For each segment $i$, we calculated the mean $\langle \Delta C/C_{\rm true} \rangle_{i}$, and for each redshift, the standard deviation of these segment means, $\sigma_{\langle \Delta C/C_{\rm true} \rangle_{i}}$.
The error on the mean at a given redshift is then given by
\begin{equation}
\sigma_{\langle \Delta C/C_{\rm true} \rangle} =
\frac{\sigma_{\langle \Delta C/C_{\rm true} \rangle_{i}}}{\sqrt{N}},
\end{equation} 
where the $N$ is the number of segments.
The redshift trend is well approximated by the best-fit power law
\begin{equation}
\label{continuum correction fit}
\Delta C/C_{\rm true}=1.58\times10^{-5}(1+z)^{5.63} 
\end{equation}
over the redshift range covered.
We use this fit to correct our $\tau_{\rm eff}$ measurement for continuum errors in \S \ref{continuum estimation}. 

Let the superscripts $true$ and $est$ refer to true values and values estimated from the continuum fitting procedure, respectively.
Then
\begin{equation}
\label{tau eff true est}
\tau_{\rm eff}^{\rm true}=-\ln{\langle F \rangle^{\rm true}},~~~\tau_{\rm eff}^{\rm est}=-\ln{\langle F \rangle^{\rm est}},
\end{equation}
with
\begin{equation}
\label{mean f true est}
\langle F \rangle^{\rm true}=
\left\langle
\frac{F_{\rm abs}}{C_{\rm true}}
\right\rangle,~~~
\langle F \rangle^{\rm est}=
\left\langle
\frac{F_{\rm abs}}{C_{\rm est}}
\right\rangle.
\end{equation}
Note that
\begin{equation}
\label{mean f true vs mean f est}
\langle F \rangle^{\rm true}=
\left\langle
\frac{F_{\rm abs}}{C_{\rm true}}
\right\rangle
=
\left\langle
\frac{F_{\rm abs}}{C_{\rm est}}
\frac{C_{\rm est}}{C_{\rm true}}
\right\rangle
=
\left\langle
\frac{F_{\rm abs}}{C_{\rm est}}
\right\rangle
\left\langle
\frac{C_{\rm est}}{C_{\rm true}}
\right\rangle
=
\langle F \rangle^{\rm est}
\left\langle
\frac{C_{\rm est}}{C_{\rm true}}
\right\rangle,
\end{equation}
where the second-to-last step holds because the local transmission can be assumed statistically independent of the continuum correction.
Combining equations \ref{tau eff true est}, \ref{mean f true est}, and \ref{mean f true vs mean f est}, we find
\begin{equation}
\label{tau eff true from tau eff est}
\tau_{\rm eff}^{\rm true} = \tau_{\rm eff}^{\rm est} - 
\ln{\left[
1 - 
\left\langle
\frac{\Delta C}{C_{\rm true}}  
\right\rangle
\right]}.
\end{equation}
The statistical error on $\langle \Delta C/C_{\rm true} \rangle$ is negligible, as can be seen from Figure \ref{continuum trend}, and so the statistical error on $\tau_{\rm eff}^{\rm true}$ is just that on $\tau_{\rm eff}^{\rm est}$.
We discuss the systematic error on our continuum correction in the next section.
\\ \\
\subsection{Uncertainty on our Continuum Correction}
\label{continuum correction uncertainty}
The continuum correction obtained above depends on the properties of the mock spectra we generated. 
In particular, it is a function of the assumed $\Gamma^{bkg}$.
The $z=4.5$ panel of Figure \ref{example cont fits}, for example, shows the mock spectra obtained by setting \Gbkg~equal to half and twice the fiducial value, for a fixed line of sight through the IGM.
We have used values of $\Gamma^{bkg}$ that were estimated from previous measurements of $\tau_{\rm eff}$ (Schaye et al. 2003\nocite{2003ApJ...596..768S} for the Bolton et al. 2005 points\nocite{2005MNRAS.357.1178B}; Fan et al. 2002\nocite{2002AJ....123.1247F} for the Meiksin \& White 2004\nocite{2004MNRAS.350.1107M} point), which were generally less precise than ours and to which no continuum corrections were applied.\footnote{\cite{2002AJ....123.1247F} extrapolated the continua of their $z\sim6$ quasars from the red side of the \Lya~emission line rather than directly estimating the continuum levels in the \Lya~forest.}
Our estimated continuum correction may therefore not be self-consistent.

To estimate by how much our continuum bias correction may be biased, fix some redshift and let $\tau^{\rm peak}_{0}$ be the characteristic optical depth of the transmission peaks (which are the points through which we fit a spline to estimate the continua; see \S \ref{continuum estimation}) for the true background photoionization rate $\Gamma^{0}$.
Let $\tau^{\rm peak}$ be the corresponding quantity for some other photoionization rate $\Gamma$.
Neglecting redshift-space distortions, $\tau^{\rm peak}=\tau^{\rm peak}_{0}(\Gamma^{0}/\Gamma)$ (e.g., Eq. \ref{gp tau}).
If we use an arbitrary value $\Gamma$ in generating our mock spectra, then the continuum level estimated from the mock spectra, $C_{\rm est}$, will differ from the correct value that would have been obtained had we assumed the true $\Gamma_{0}$, $C_{\rm est,0}$:
\begin{equation}
\label{C est over C est 0}
\left\langle
\frac{C_{\rm est}}{C_{\rm est,0}}
\right\rangle
\approx
\left.
\frac{F^{\rm est}}{F^{\rm est,0}}
\right|_{\rm peak}
\equiv
\frac{e^{-\tau^{\rm peak}}}{e^{-\tau^{\rm peak}_{0}}}
=
\exp{[-\tau^{\rm peak}_{0}(\Gamma_{0}/\Gamma - 1)]}
\approx
1 -\tau^{\rm peak}_{0}(\Gamma_{0}/\Gamma - 1),
\end{equation}
where the last step is valid if the argument of the exponential is $\ll 1$.
Define $\Delta \tau_{\rm eff}\equiv \tau_{\rm eff}^{\rm true}(\Gamma_{0})-\tau_{\rm eff}^{\rm true}(\Gamma)$, where the argument of $\tau_{\rm eff}$ indicates the background photoionization rate that is assumed in correcting for continuum bias.
Substituting Equation (\ref{tau eff true from tau eff est}), simplifying, and using Equation (\ref{C est over C est 0}) then gives
\begin{equation}
\frac{\Delta \tau_{\rm eff}}
{\tau_{\rm eff}^{\rm true}} =
\left.
\ln{
\left[
\frac{\langle C_{\rm est}/C_{\rm est,0} \rangle(\Gamma)}
{\langle C_{\rm est}/C_{\rm est,0} \rangle(\Gamma_{0})}
\right]
}
\right/
\tau_{\rm eff}^{\rm true}
=
\frac{
\ln{[1 - \tau_{0}^{\rm peak}(\Gamma/\Gamma_{0} - 1)]}
}
{
\tau_{\rm eff}^{\rm true}
}
\approx
\frac{
\tau_{0}^{\rm peak}
}
{
\tau_{\rm eff}^{\rm true}
}
(1 - \Gamma/\Gamma_{0})
.
\end{equation}

An idea of the magnitude of this fractional error is obtained by assuming that the continuum correction we have calculated using values of $\Gamma^{bkg}$ from the literature is close to exact.
Since the estimated continua pass through the transmission peaks, $\langle \Delta C/C_{\rm true} \rangle \approx \tau_{0}^{\rm peak}$.
Using the fits in Equations (\ref{tau eff power law fit}) and (\ref{continuum correction fit}) for $\tau_{\rm eff}^{\rm true}$ and $\langle \Delta C/C_{\rm true} \rangle$, respectively,
\begin{equation}
\frac{\Delta \tau_{\rm eff}}
{\tau_{\rm eff}^{\rm true}}
\approx
0.0088(1+z)^{1.71}
\left[
1 - 
\frac{\Gamma}
{\Gamma_{0}}(z)
\right].
\end{equation}

\cite{2005MNRAS.357.1178B} performed a detailed analysis of the sources of uncertainty in their determination of $\Gamma^{bkg}$ from the flux decrement method, concluding that it is known within $\approx50 \%$ at $2\leq z \leq4$, giving
\begin{equation}
\frac{|\Delta \tau_{\rm eff}|}
{\tau_{\rm eff}^{\rm true}}
\lesssim
0.0044(1+z)^{1.71}.
\end{equation}
Thus, our continuum-corrected $\tau_{\rm eff}$ measurement may be off by as much as $\approx(3, 5, 7)\%$ at $z=(2, 3, 4)$.\footnote{The \cite{2005MNRAS.357.1178B} uncertainty estimates do not include the systematic continuum bias. However, their total uncertainties of order $50\%$ are much larger than the effect of continuum bias alone.}
Note, however, that the continuum correction is expected to be a smooth function of redshift, so that uncertainties on it are unlikely to affect any conclusion that we may draw about the presence of a narrow feature in the redshift evolution of $\tau_{\rm eff}$.

The continuum correction $\langle \Delta C/C_{\rm true} \rangle \approx \tau_{0}^{\rm peak}$ also depends on the cosmological parameters through the expression for the optical depth,
\begin{equation}
\label{gp tau appendix}
\tau=
\frac{\pi e^{2} f_{\Lya}}{m_{e} \nu_{\Lya}}
\frac{1}{H(z)}
\frac{R(T)n_{\rm HII}n_{e}}{\Gamma^{bkg}}
\propto
\frac{\Omega_{b}^{2}}
{H_{0}\sqrt{\Omega_{m}}},
\end{equation}
where the variables are defined as in equation \ref{gp tau appendix}, and we have neglected the cosmological constant at the redshifts $z\gtrsim2$ of interest.
However, these cosmological parameters are known much more precisely than \Gbkg~\citep[e.g.,][]{2007ApJS..170..377S}.
We see from equation \ref{gp tau appendix} that $\tau_{0}^{\rm peak}$ also depends on the temperature $T$ of the IGM, but this dependence is degenerate with that on \Gbkg.
The uncertainties on \Gbkg~quantified by \cite{2005MNRAS.357.1178B} include a contribution from the uncertainty on $T$.
The degeneracy therefore implies that the error estimates above on our continuum-corrected measurement are actually overestimates.

A more accurate and self-consistent determination of the continuum correction can in principle be achieved by iteratively re-estimating it, at each step inferring \Gbkg~from the data using the current correction, and using it for the next iteration until convergence is attained. 
We defer this procedure to future work, in which we will accurately infer \Gbkg.
We here simply emphasize the uncertainty, estimated above, on continuum correction.

\bibliography{references}

\end{document}